\documentclass{acmtrans2e}

\usepackage{alltt}
\usepackage[]{fontenc}
\usepackage{amssymb}
\usepackage{amsmath}
\usepackage{moreverb}
\usepackage{umlaut}
\usepackage{pslatex}
\usepackage[dvips]{epsfig}
\usepackage{color}
\usepackage{comment}
\usepackage{listings}

\newtheorem{define}{Definition}
\newtheorem{theorem}{Theorem}
\newtheorem{lemma}{Lemma}

\begin{document}

%==============================================================================

\markboth{Daniel Pfeifer and Peter C. Lockemann}{Theory and Practice of Transactional Method Caching}
\runningfoot{}

\title{Theory and Practice of Transactional Method Caching}

\author{Daniel Pfeifer and Peter C. Lockemann\\
Institute for Program Structures
 and Data Organisation (IPD)\\
 Universität Karlsruhe, Germany\\
 \{pfeifer, lockeman\}@ipd.uni-karlsruhe.de\\
 \today}

\begin{abstract}
Nowadays, tiered architectures are widely accepted for
constructing large scale information systems. In this context
application servers often form the bottleneck for a system's
efficiency. An application server exposes an object oriented
interface consisting of set of methods which are accessed by
potentially remote clients. The idea of \emph{method caching} is
to store results of read-only method invocations with respect to
the application server's interface on the client side. If the
client invokes the same method with the same arguments again, the
corresponding result can be taken from the cache without
contacting the server. It has been shown that this approach can
considerably improve a real world system's efficiency.

This paper extends the concept of method caching by addressing the
case where clients wrap related method invocations in ACID
transactions. Demarcating sequences of method calls in this way is
supported by many important application server standards. In this
context the paper presents an architecture, a theory and an
efficient protocol for maintaining full transactional consistency
and in particular serializability when using a method cache on the
client side. In order to create a protocol for scheduling cached
method results, the paper extends a classical transaction
formalism. Based on this extension, a recovery protocol and an
optimistic serializability protocol are derived. The latter one
differs from traditional transactional cache protocols in many
essential ways. An efficiency experiment validates the approach:
Using the cache a system's performance and scalability are
considerably improved.

%An important component to guarantee serializability is a special scheduler that coordinates
%the use cached method calls with the transactional resources that underly the application server.
%The related scheduler can cooperate with ordinary is located outside a transactional resource, e.g. a relational database
%between ordinary is located on the server side.
\end{abstract}

%\begin{keywords}
%H.2.4.o Transaction Processing,
%H.3.4.b Distributed Systems,
%C.4 Optimization
%\end{keywords}

\category{H.2.4.o}{Information Systems}{Database Management}[Systems, Transaction Processing]
\category{H.3.4.b}{Information Systems}{Information Storage and Retrieval}[System and Software, Distributed Systems]

\maketitle

\section{Introduction}
\label{introduction}

Modern large-scale client-server-based information systems follow a tiered
architecture. The most common solution is the three-tier
architecture consisting of a presentation tier, an application tier and a data tier.
E.g. for a typical web application, a servlet-enabled web server implements the presentation tier and
a (relational) database system implements the data tier. Application server technologies
such as EJB \cite{ejb} or corresponding parts of the .NET Framework \cite{dotnet}
are often used to realize the application tier. They offer an object oriented interface
consisting of a set of service methods to their clients, the so called \emph{service interface}.
In order to centralize business logic but also for better system scalability,
the different tiers are usually hosted on separate machines in a local network. This makes invoking a
service method a costly affair, since it requires a remote method call which passes the application server's
infrastructure and often incurs database accesses.

Consequently, application servers tend to become the bottleneck of an information system
in respect to its performance and scalability.
Many solutions have been proposed to tackle this problem including
dynamic web caching \cite{oracle_reverse_proxy,challenger99scalable,cache_portal_2},
method caching \cite{pfeifer2},
application data caching \cite{jcache},
database caching \cite{oracledbcache,db2,timesten_2} and special design patterns \cite{ejb_patterns}.
%Again there are many different concepts to provide weak or strong consistency.

We concentrate on \emph{method caching} whereby \emph{results of service method calls are cached on the client side of
an application server}. E.g. in case of a tiered web application, an application server's client is usually a servlet-enabled web server.
Alternatively, an application server's client could also be an end-user program with rich a graphical user interface.

If the client code invokes a service method that does not have any
side effects, its result may be cached for later reuse on the client side.
If the client code calls the same method with the same arguments again, the result can be
read from the cache without contacting the application server.

\cite{pfeifer2} showed that this
approach can be pursued transparently, so that usually neither the client-side nor the server-side application code
has to be aware of a related cache's presence. Moreover, it validated
that a method cache can considerably improve performance and scalability of real world applications.

For caching approaches the most challenging part is usually to guarantee cache consistency.
\cite{pfeifer2} also demonstrates how strong cache consistency
can be asserted for the price of added efforts on the part of an application developer
who has to describe certain interdependencies between methods.
However strong cache consistency does not cover the case where service method calls
are demarcated by client-side transactions.

Consequently, this paper extends the idea of method caching by addressing the case where
the client code wraps service method invocations in ACID transactions. This type of transactions
is explicitly supported by popular
application server technologies such as EJB and .NET. This paper presents an architecture and a theory that enables
\emph{transactional caching of method results on the client side} while maintaining complete
transactional consistency and in particular serializability.
Moreover, we discuss how to preserve important recovery properties when using a transactional method cache.

In this context many important assumptions differ from the ones
that govern conventional transactional cache protocols such as presented in \cite{franklin97transactional}.
In particular, we do not assume that a protocol for transactional method caching
can be tightly integrated into the database system that underlies the application server.
In practice, such an expectation would be unrealistic because commercial database systems
do not allow a deep engagement in their internal transaction manager.
Instead we propose an independent component, called the $m$-scheduler, for scheduling
cached method results while asserting full transactional consistency. The $m$-scheduler
is located in between the application server and the underlying database system,
cooperates with a transactional method cache on the client-side and
makes conservative assumptions about the database system's transaction management.

The remainder of this paper is organized as follows:
First we clarify the scope to which transactional method caching may be applied
and explain how an application server architecture should be extended to enable this caching approach (Section \ref{architecture}).
In order to build an $m$-scheduler and a related cache protocol,
it is useful to extend the conventional notion of transactions. Section \ref{theory} develops a
theory for transactional method caching on the basis of the classical 1-version and multiversion transaction formalisms.
Using this theory, Section \ref{protocol} develops a serializability protocol for
scheduling cache hits for cached method results inside transactions.
The protocol is optimistic but differs from existing transactional cache protocols such as OCC \cite{223787} in many essential ways.
Before, Section \ref{recoverysec} discusses how conventional recovery qualities can be
assured in the presence of a transactional method cache.
To demonstrate that the approach pays off, the paper presents
an efficiency experiment for an EJB-based application server system (Section \ref{evaluation}).
Section \ref{relatedwork} outlines the relationships between our contribution and existing caching approaches for web applications
as well as existing transaction protocols. We conclude with a summary and prospects to future work.

\section{General Architecture}
\label{architecture}

\subsection{Client-Side Transactions for Application Servers}

\begin{figure}[!tb]
\centering
\includegraphics[height=6cm]{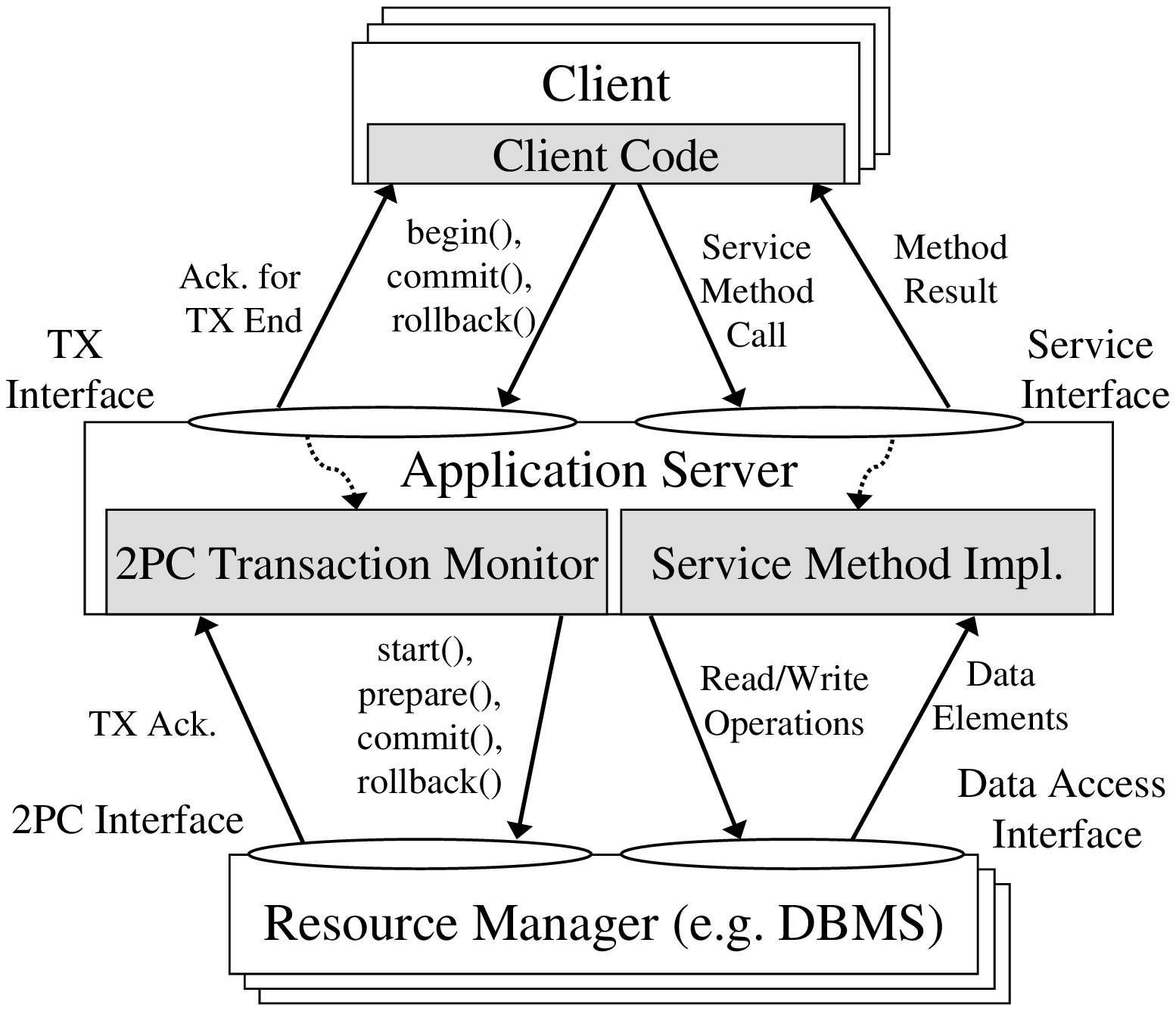}
\caption{Architecture of an Application Server Supporting Client-Side Transactions}
\label{txarch}
\end{figure}

%Before explaining how and where transactional method caching can be applied,
This section highlights the general concept of client-side application
transactions and the respective infrastructure.
Figure \ref{txarch} illustrates an architecture enabling client-side
transactions in conjunction with service interfaces: An application server offers
two interfaces, the service interface and the transaction interface. Both interfaces
can be used via remote method calls from a client. E.g. for EJB, the service interface
technically consists of a set EJB Home and EJB Remote Interfaces (which are Java interfaces) while
the transaction interface adheres to the Java Transaction API \cite{jta}.\footnote{
Note that the term "service interface" abstracts from the actual number of programming
language interfaces for an application server standard.}
Using these interfaces, a client can wrap a sequence of service method invocations in an ACID transaction.
The application server executes the client's
service method invocations and relies on one or more transactional resources (e.g. databases)
to enable transactional consistency. To achieve this, the application server state (as far as relevant
to clients) is derived from the state of the transactional resources. If a transactional resource
is a relational database, this is typically realized by SQL statements inside service method
implementations or by object relational mappings between application server objects and database table rows.
As shown in Figure \ref{txarch} a service method implementation may therefore read and write data elements
via the data access interface of the underlying database system.

For every transaction that a client begins, the application server starts
a transaction on every registered resource manager (e.g. a database transaction) and keeps it open for as long
as the client transaction is open. All service method invocations inside a client's
transaction are tied to a respective resource transaction for every participating resource manager.
To realize this, the resource managers are usually expected to provide a transaction demarcation interface
according to the XA standard \cite{opendtp}. When committing a client transaction, the application
server acts as a transaction monitor and commits all
respective resource transactions using a two-phase commit protocol.
Due to this mechanism, transactional qualities of resource transactions are more or less inherited by
client-side transactions. E.g., if there is only one participating resource manager and it guarantees serializability
then the client transaction will also be serializable.

Note that typically, application servers do not guarantee \emph{global} serializability across multiple resource managers but only
ascertain local serializability and atomic commits. The approach of this paper does not try change this fact but offers the same
degree of consistency in the presence of client-side method caches. Therefore, the actual number of
resource managers is mostly irrelevant to this contribution (given that there is at least one such entity).

\lstset{numbers=left,numberstyle=\fontsize{8}{8}\selectfont,
basicstyle=\fontsize{8}{9}\selectfont\ttfamily,
commentstyle=\fontsize{8}{9}\selectfont\rmfamily,morekeywords=[1]{each},
columns=fullflexible,keywordstyle=,language=Java,texcl=true,showstringspaces=false}
\begin{figure}[!tb]
\begin{lstlisting}
...
Context ctx = new InitialContext();
// Request an application transaction
UserTransaction utx = (UserTransaction) ctx.lookup("java:comp/UserTransaction");
utx.begin(); // Begin the transaction
Item item = itemSession.findItemById(20); // Invoke service methods as part of the current transaction
if (!item.price > 42) {
  item.price = 42;
  itemSession.updateItem(item);
}
utx.commit(); // Commit the transaction
...
\end{lstlisting}
\caption{Example Code of a Client-Side Transaction Using EJB}
\label{clienttxsample}
%\vspace{-0.5cm}
\end{figure}

Figure \ref{clienttxsample} presents an example of EJB-related code for a client-side transaction including service methods calls.

%In Line 4 the client requests a transaction from a central directory service.
%Then it starts the transaction (Line 7), performs some service method calls (Lines 9 to 12) and tries to commit (Line 15).

\subsection{Integrating a Transactional Method Cache}
\label{integration}

% How does transactional method caching fit into this picture?
This section explains how a transactional method cache can be integrated in the above architecture.
It shows how a service method invocation is generally processed in the presence of
a method cache and describes a base protocol for keeping the cache contents up-to-date.

\subsubsection{Base Architecture}

\begin{figure}[!tb]
\centering
\includegraphics[height=7.5cm]{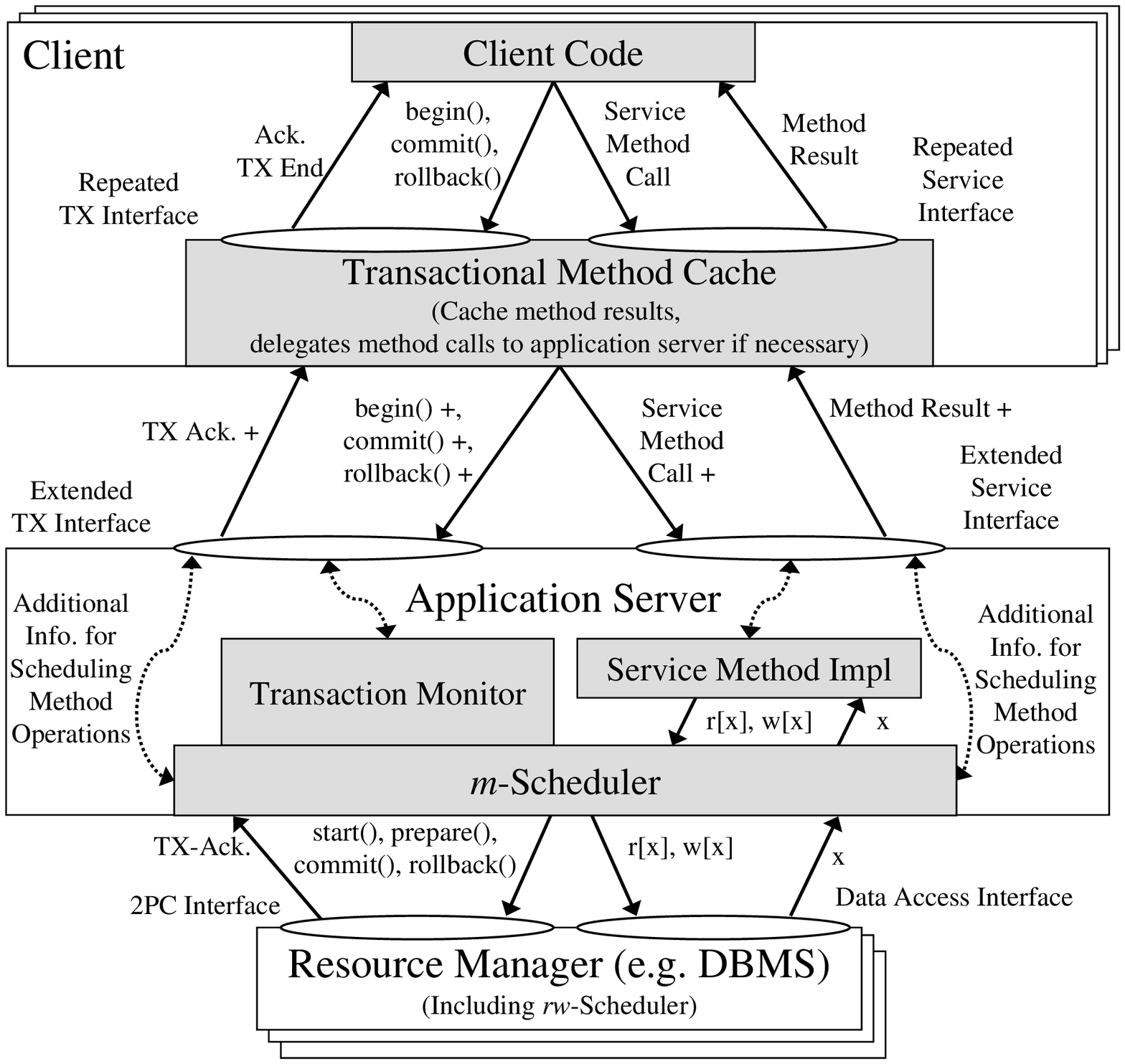}
\caption{Architecture of an Application Including a Transactional Method Cache and an $m$-Scheduler}
\label{txmarchext}
\end{figure}

Figure \ref{txmarchext} extends Figure \ref{txarch} by the components additionally
needed for transactional method caching. As described in Section \ref{introduction} the cache
is located at the client and implements the application server's transaction interface as
well as its service interface.\footnote{Technically this can by realized by applying the
the design pattern "dynamic proxy" (\cite{dynamicproxy}) or by generating the respective classes statically \cite{pfeifer2}.}

For the client code, the presence of the cache is completely transparent -- it performs its method calls as usual.
However, service method invocations and calls to demarcate transactions are now intercepted by the transactional method cache.
For every service method call the cache checks if a related method result is in its store. If so, it returns the result
to the client right away. Otherwise it delegates the call to the server where it is (almost) executed as usual.
The cache always forwards calls for demarcating client-side transactions to server.

In order to exchange additional cache consistency information, all remote method invocations might transfer extra data.
This is indicated in Figure \ref{txmarchext} by a plus sign added to a respective label.
(Most modern remote method invocation protocols allow for these kind of extensions.)
When a method call arrives at the server, the additional information from the method cache is passed on to the $m$-scheduler.
As soon as the call's result is about to be returned to the client, the $m$-scheduler attaches consistency information
which will be processed by the cache.

The approach leaves the conventional message flow between client and
server intact, since additional data is always piggy-backed to ordinary remote method calls.
Only in case of cache hits, the information flow changes since client server communication is avoided.
This lazy way of exchanging cache consistency information keeps the communication cost at
a minimum but requires transactional method cache protocols that are optimistic.

\subsubsection{Base Protocol}
\label{baseprot}

The following paragraphs describe the \emph{base protocol} for transactional method caching.
Note that this protocol does not yet guarantee serializability. It merely asserts that
cache content is created for read-only method invocations
and that stale cached method-results will be invalidated soon after
a respective write operation. In later sections we will see how the base protocol can be
extended to ascertain transactional consistency.

Also note, that the base protocol as described next refers to just one client cache
whereby the corresponding client might run several concurrent transactions.
However, the protocol can easily be extended to function to with multiple clients. (The details are
omitted in favour of a compact presentation.)

Let $m$ be service method and $o.m(a)$ be a corresponding method invocation
comprising the this-object $o$ and the argument list $a$.
When $o.m(a)$ reaches the cache it checks if
the result for the cache key $(m,o,a)$ is in its store.
For a cache hit, the result is returned to the client code straight away.
Further, for every active local client transaction $T_i$ the cache keeps an initially empty list $L_i$
and enters into it all method results that were returned from the cache on behalf of transaction $T_i$.
In order to do so, every cached method result is assigned a unique identifier which is entered in $L_i$.

If a cache miss occurs in $T_i$ or if the client tries to commit $T_i$, the respective
method call is delegated to the application server.
The method cache attaches the list $L_i$ of the respective transaction $T_i$
to the call and sends it to the server.
On the server side, the call is executed as usual, however $L_i$ is
forwarded to a new component -- the so called \emph{$m$-scheduler}.
The $m$-scheduler is in charge of scheduling the use of cached method results in such a
way that a client transaction $T_i$ remains consistent, i. e., serializable.
It can do so because it knows all cache hits of $T_i$ from the
respective list $L_i$, and also, it observes all data access operations that
service method implementations perform via resource managers.

Take a cache miss so that the method call $o.m(a)$ from above might cause several
read and write operations on a relational database. The $m$-scheduler observes these operations,
keeps track of them in an operation list $l_i$, and passes the operations on to the database system.
For now we assume that $l_i$ consists of operations of the type $r[x]$ and $w[x]$ with $x$ representing
a data element of the database. However, as it will be discussed later, there are challenges in
identifying data elements such as $x$.

When the execution of $o.m(a)$ finishes at the server, the $m$-scheduler checks if there are any write operations
in the operation list $l_i$. If not, the respective method invocation left the database state unchanged and
will become a candidate for caching.
In this case the $m$-scheduler associates a globally unique identifier $(i,k)$ with $o.m(a)$ where
$i$ represents the transaction $T_i$ in which $o.m(a)$ was computed and $k$ identifies
$o.m(a)$ inside $T_i$. Moreover, the $m$-scheduler maintains
a global table $V$ to associate all identifiers of cached method calls (from all transactions)
with all data elements that were read during a respective method execution. So,
for $o.m(a)$ it will enter $(i,k)$ and the respective data elements (such as known from $l_i$) in $V$.
When the application server sends the result $r$ from $o.m(a)$'s execution to the client, the respective
message also contains the tuple $(i,k)$. This tells the cache that $r$ should be cached and it saves
both $r$ and $(i,k)$ together with the cache key $(m,o,a)$ in its store.

If, on the other hand, $o.m(a)$ did cause one or more write operations, the system behaves differently:
Let $x$ be a data element which was written on behalf of $o.m(a)$.
Using $V$ the $m$-scheduler determines all identifiers of cached method results at whose computation
$x$ was read and collects them in an invalidation list $h$.
% No for OCTP: Afterwards the respective entries are removed from $V$.
The server attaches $h$ to the result message which contains $r$ and sends it to the client.
When the client receives the message it removes all method results from the cache which are
identified by elements in $h$. Eventually it returns $r$ to the client code.

To sum up, the $m$-scheduler needs identifiers for cached method
calls like $(i,k)$, lists like $L_i$, $l_i$ and $h$ as well as the
table $V$ to enable consistent transaction executions and to keep
the cache up-to-date. Using $L_i$ the $m$-scheduler gets to know
what cached method calls were accessed in a transaction. Using $V$
the $m$-scheduler can tell what data elements were read to produce
cached method results and also it can derive what cached method
results must be invalidated. Using $(i,k)$ the $m$-scheduler can
associate cache hits with entries from $V$.

\lstset{numbers=right,numberstyle=\fontsize{8}{8}\selectfont,
basicstyle=\fontsize{8}{9}\selectfont\ttfamily,
commentstyle=\fontsize{8}{9}\selectfont\rmfamily,morekeywords=[1]{each},
columns=fullflexible,keywordstyle=,language=Java,texcl=true,showstringspaces=false}
\begin{figure}[!tb]
\begin{lstlisting}[mathescape]
interface DE {} // Representation of a data element (just a marker interface)
class MId { int k,l; } // ID of a cached method result
class Op { boolean read; DE x; } // Representation of a database operation $r[x]$ or $w[x]$
class T { // Representation of a transaction $T_i$
  int id; // Transaction ID
  List<Op> l = $\emptyset$; // Database operations for the current method execution ($l_i$)
  int nextMId = 0; // Counter for new IDs of cached method results
  ...
}
class Req { // A service method call which is forwarded to the server
  int txId; // ID of the transaction containing the call
  Object o; Method m; Object[] args; // Method call details
  // Recent client-side cache hits for the given transaction ($L_i$)
  List<MId> L;
}
class Res { // Response for a service method execution
  Object r; // The execution's result
  boolean cachable; // Whether the result is cachable or not
  MId m = null; // If result is cachable: the ID of the result
  List<MId> h; // IDs of recently invalidated cached method results
}

class MScheduler {  // Representation of the $m$-scheduler
  Rel<DE,MId> V = $\emptyset$;   // Relates $x$ with IDs of cached method results
  Map<int,T> txId2T = $\emptyset$; // Relates a transaction's ID with its transaction object

  void handleRequest(Req req) {   // $m$-scheduler part for handling a request of a service method execution
    for each m $\in$ req.L // Iterate over all recent cache hits of the considered transaction and schedule them
      methodOp(txId2T(req.txId), m); // (for details see later)
  }
  void completeResponse(Res res, T t) {   // Complete the response of a service method execution
    res.cachable = true; // At first, assume that the result is cachable
    for each op $\in$ t.l
      if (!op.read) { // If the method executed a write operations, $\ldots$
        res.cachable = false; // $\ldots$ it is not cachable
        // Update $h$ to invalidate the respective cache entries at the client
        for each m $\in$ V(x) res.h.add(m);
      }
    if (res.cachable) { // If the result will be cached, $\ldots$
      res.m = new MId(t.id, t.nextMId++); // $\ldots$ generate its ID and $\ldots$
      for each op $\in$ l // $\ldots$ register it at the server using $V$
        V.put(op.x, res.m);
    }
    l.clear(); // Clear the database operations list for the next method execution
  }
  ...
}
\end{lstlisting}
\caption{Java Pseudo Code for the Base Protocol's Aspects at the $m$-Scheduler}
\label{baseprotcode}
\end{figure}

Figure \ref{baseprotcode} illustrates the base protocol's data structures and some of its important
implementation aspects at the server side.\footnote{
Note that in order to represent data types conveniently, the code applies
parametric polymorphism (also known as "generics" in the Java world \cite{generics}).
E.g., the polymorphic type \texttt{Rel<A,B>} stands for finite
relations $R \subseteq$ \texttt{A}$\times$\texttt{B} and the type
\texttt{Map<A,B>} represents finite functions \texttt{A}$\rightarrow$\texttt{B}.}
The classes \texttt{Req} and \texttt{Res} represent
the requests and the responses of service method calls addressing the server.
The classes' field names match the names used in the protocol description from above.

At the server side, the $m$-scheduler drives the base protocol in respect to handling requests
and generating responses. In this context the application server is supposed to call \texttt{MScheduler.handleRequest()}
when it receives a remote service method call.
After the application server has computed the method call's result $r$, it invokes \texttt{completeResponse()}.
This way the $m$-scheduler can add all missing base protocol information to the response object.
Eventually, the server sends the completed response object to the client.

\subsubsection{Integrated Transaction Scheduling}

Although method caching happens on the client side, cache consistency
is provided by the $m$-scheduler (on the server side).
Without a transactional method cache in place, client transactions are mainly based
on the transaction management of resource managers. For this purpose, every
resource manager has its own unit for scheduling transaction operations, the so called \emph{$rw$-scheduler}.
E.g. the $rw$-scheduler applies a serializability protocol such as two-phase locking \cite{BHG87}, 2-version two-phase locking
or FOCC \cite{2041}. Unfortunately, the use of cached method results is beyond an $rw$-scheduler's control
but still affects transactional consistency. Therefore, the $m$-scheduler and a respective $rw$-scheduler
must cooperate in order to provide consistent client transactions.

Since resource manager products such as relational database management systems\\
(RDBMs) cannot be easily prepared for such
an integration, we propose a layered approach for scheduling transactions in the presence of a method cache.
Using this approach the resource manager is completely unaware of an $m$-scheduler
and performs its tasks as usual.

The $m$-scheduler intercepts all transaction operations that address the resource
manager and on top of it, it schedules the use of cached method results.
In order to do so, it makes conservative assumptions about the $rw$-scheduler's behavior and handles conflicts
resulting from the use of cached method results and conventional write operations.
Using the data structures from above it has all information at hand to perform this task.
The next part of this paper is devoted to developing a theory for how an $m$-scheduler can
produce serializable transactions under these conditions.
The general idea of separating different parts of a transaction scheduling process along
certain types of data operations can be found in \cite{BHG87}. We build on this idea for creating an
integrated scheduler consisting of an $m$-scheduler and an $rw$-scheduler.

Note that it is a crucial requirement for the $m$-scheduler to
observe \emph{all} transaction operations addressing the resource manager. Otherwise,
it might miss potential conflicts between operations and therefore generate non-serializable histories.

As mentioned above, there is an additional challenge when constructing an $m$-scheduler because
it has to observe access operations in respect to single data elements from a database.
E.g. if the $m$-scheduler should integrate with an RDBMS, database elements might be
table rows. Since the $m$-scheduler acts outside of the RDBMS,
it can only observe database access on the basis of SQL statements. Unfortunately
SQL statements specify data elements only descriptively and so the $m$-scheduler is unable to
directly identify data elements as needed. As a rather pragmatic solution to this problem, we expect an
application developer to help out by providing the necessary information via some extra
code inside service method implementations.
The extra code is inserted after a corresponding SQL statement
and refers to the $m$-scheduler in order to tell it what data elements the SQL statement accessed.
It is up to the application developer to find a useful representation for identifying data elements.
From our experience, key values of table rows are mostly a good choice.

% The issues related to observing access to data elements can be entirely avoided if the $m$-scheduler
% forms an integral part of the resource manager. The main contribution of this paper, namely a theory
% for scheduling cache can still be applied to this case.

% \begin{figure}[!tb]
% \centering
% \includegraphics[height=3.4cm]{schedulerinteg.eps}
% \caption{Abstract Interfaces for Integration of an $m$-Scheduler and $rw$-Scheduler}
% \label{txmcache_uml}
% \end{figure}

\section{Transaction Theory for Method-Based Caching}
\label{theory}

\subsection{MC-Transactions and MC-Histories}

In order to produce serializable histories in conjunction
with method caching, one has to represent the use of cached method results
in transaction histories. This section extends the  notion
of conventional transactions and 1-version histories such as presented in \cite{BHG87}
by introducing a new operation
that indicates the use of a cached method result inside a transaction.
As opposed to conventional read and write operations we call such an operation
a \emph{method operation}.

A benefit of method operations is that they accurately and naturally represent
of the use of cached method results in a transaction formalism. More importantly, they
enable the development and the verification of non-trivial serialization protocols
for $m$-schedulers. One such protocol will be described in Section \ref{protocol}.

For an intuitive understanding of method operations we take a look at a corresponding
history before we come up with a proper definition for it.
Consider the following history:
\[H_1 = r_1^4[y]r_1^4[x]c_1 w_2[x]c_2 m_3^{1,4}r_3^5[x]c_3.\]
How does it differ from a conventional 1-version history?
First of all, we have read operations with superscripts such as $r_1^4[x]$.
This operation is just like an ordinary 1-version read operation (e.g. like $r_1[x]$) except
that the superscript $4$ is an identifier for the method call on whose behalf
the read operation was performed. The respective method call is executed on the server side
and so it produces ordinary read operations at the resource manager.
As the method call reads two data elements, there is a series of read
operations with the same superscript, namely $r_1^4[y]$ and $r_1^4[x]$.
Since the method call with the ID $4$ in $T_1$ only reads data, its result may be
cached on the client side. Afterwards it is available for cache hits (which might
occur in other transactions). Note that from a technical point of view, the superscripts
for read operations are created and used by the $m$-scheduler. They are not visible
and not relevant to a resource manager's $rw$-scheduler.

Secondly, $H_1$ contains the method operation $m_3^{1,4}$. \emph{It reflects an access to a
cached method result} in transaction $T_3$. The index $3$ specifies that $m_3^{1,4}$ belongs to $T_3$.
Furthermore, the superscript of $m_3^{1,4}$ uniquely identifies the cached method
result to which it refers: It is just the result that was produced
by the operations $r_1^4[y]$ and $r_1^4[x]$
of $T_1$. So the number $1$ in the superscript of $m_3^{1,4}$ refers to $T_1$
and the number $4$ identifies the method call with the ID $4$.

We have just covered the most relevant aspects of MC-histories and how they extend conventional
1-version histories. The following definitions implement these ideas.

\begin{define}
\label{transaction}
An \emph{MC-transaction} $T_i$ is a set of operations with a partial
ordering relation $<_i$, where
\begin{itemize}
\item
\ $T_i \subseteq \{ w_i[x], r_i^j[x], m_i^{k,l}\ |\ x \textrm{\ is a data element\ } \wedge j,k,l \in \mathbb{N} \setminus\{0\} \} \cup \{ a_i, c_i \}$,
\item
\ $a_i \in T_i \Leftrightarrow c_i \notin T_i$,
\item
\ $\forall p \in T_i: p \notin \{ a_i, c_i\} \Rightarrow (p <_i a_i \vee p <_i c_i)$,
\item
\ $\forall r_i^j[x], w_i[x] \in T_i: r_i^j[x] <_i w_i[x] \Leftrightarrow \neg (w_i[x] <_i r_i^j[x])$.
\end{itemize}
\end{define}

Besides introducing method operations, MC-transactions require every read operation to have a superscript.
Note that a read operation's superscript is only necessary to "reference it" from method operations
as explained for the history $H_1$.\footnote{Technically, superscripts for read operations form an
extension of conventional 1-version transactions because a respective
transaction may contain several read operations of the same data element whereas this is
not the case for a transaction such as defined in \cite{BHG87}. However, this detail has no
major impact on transaction theory.}

\begin{define}
\label{mchist}
Let $\{T_1, \ldots, T_n\}$ be a set of MC-transactions. An \emph{MC-history} $H$ is defined as
$H =  \bigcup_{i = 1}^n T_i$ with a partial ordering relation $< \supseteq \bigcup_{i = 1}^n <_i$.
Furthermore, the following condition must hold:
\[\forall m_i^{k,l} \in H: k \in \{1, \ldots, n \} \wedge \forall r_k^l[x] \in H: r_k^l[x] < m_i^{k,l}.\]
\end{define}

The last condition of Definition \ref{mchist} ensures that every
$m_j^{k,l}$ refers to a $T_k$, that exists in $H$. However,
it is not necessary there exist any read operations of the form $r_k^l[\ldots]$ in $H$.

\begin{define}
\label{conflict}
The function $d(p)$ returns the set of data elements of an operation $p$ in an MC-history $H$ as follows:
\[d(r_i^j[x]) = d(w_i[x]) = \{ x \}, d(m_i^{k,l}) = \{x\ |\ \exists r_k^l[x] \in H\ \}.\]
Further, $a(p)$ shall be the type of an operation $p \in H$, so $a(r_i^j[x]) = r$, $a(w_i[x]) = w$
and $a(m_i^{k,l}) = m$.

Two operations $u_i, v_j\in H$ \emph{conflict with each other}, expressed by $u_i \nparallel v_j$, iff
\begin{gather*}
d(u_i) \cap d(v_j) \neq \emptyset \wedge \Big(\big(T_i \neq T_j \wedge (a(u_i) = w \vee a(v_j) = w)\big) \vee\\
\big(a(u_i) = w \wedge a(v_j) = m\big) \vee \big(a(u_i) = m \wedge a(v_j) = w\big)\Big).
\end{gather*}
\end{define}

Obviously, the data elements that cause conflicts in respect to
a method operation $m_i^{k,l}$ are just the ones which are read by operations of the form $r_k^l[\ldots]$.
Consider the MC-history $H_1$ from above. It holds the following conflicts (and no others):
\[r_1^4[x] \nparallel w_2[x], w_2[x] \nparallel r_3^5[x], m_3^{1,4} \nparallel w_2[x].\]
Definition \ref{conflict} states that conflicts inside a single transaction $T_i$ are possible if one
of the conflicting operations is a write operation and the other one is a method operation.
To see why this is useful, consider the history
\[H_2 = r_1^1[x] c_1 w_2[x] m_2^{1,1} c_2.\]
Here, $w_2[x] \nparallel m_2^{1,1}$ is reasonable because $m_2^{1,1}$ refers to an $x$-value that was
read before $w_2[x]$ is performed.

As is common for conventional 1-version histories, we want to avoid MC-histories with unordered but conflicting
operations. The next definition limits MC-histories in this respect.

\begin{define}
An MC-history $H$ is \emph{well defined}, iff
\[\forall p,q \in H^{MC}: p \nparallel q \Rightarrow p < q \vee q < p.\]
\end{define}

For the rest of this paper we are only interested in well defined MC-histories. \emph{So from now on,
whenever we refer to the term "MC-history" we actually mean "well defined MC-history".}

% \qqcomment{Motivate $rw$-projection here...}

\begin{define}
\label{rwproj}
The \emph{$rw$-projection} $RW$ maps an MC-history $H$ to a history $RW(H)$ with all operations from $H$ but its
method operations, so $RW(H) = \{ p \in H\ |\ a(p) \neq m \}$.
Furthermore, it keeps all ordering relations from $H$, but those in which method operations are involved.

If $RW(H) = H$ holds for an MC-history $H$, it is called an \emph{$rw$-history}.
Similarly, if a transaction does not contain any method operations it is called an \emph{$rw$-transaction}.
\end{define}

As an example of an $rw$-projection consider
\[RW(H_1) = r_1^4[y] r_1^4[x] c_1 w_2[x] c_2 r_3^5[x] c_3.\]
Apart from the superscript of read operations $rw$-histories represent conventional 1-version histories.
Later, $rw$-projections will help us to formalize how an $m$-scheduler and $rw$-scheduler split their work
for producing an integrated schedule. Note that the $rw$-scheduler only gets to see the $rw$-projection of an MC-history.
This means that formal qualities that the $rw$-scheduler should assert, may be associated with
an $rw$-projection but not an entire MC-history.

\subsection{Multiversion Histories}

This section briefly defines a slight adaption of
conventional multiversion histories and
multiversion serializability graphs. The adaption is necessary for
a sound introduction of serializable MC-histories
which follows in Section \ref{interpret}.

\begin{define}
\label{multversionhist}
Let $\{T_1, \ldots, T_n\}$ be a set of $rw$-transactions. A \emph{multiversion history} $H$ is defined as
$H =  \{\ h(p)\ |\ p \in \bigcup_{i = 1}^n T_i)\ \}$ with a partial ordering relation $<$.
Further, the function $h$ must fulfill the following criteria:
\begin{itemize}
\item
\ $\forall a_i,c_i,w_i[x] \in \bigcup_{k = 1}^n T_k: h(a_i) = a_i \wedge h(c_i) = c_i \wedge h(w_i[x]) = w_i[x_i]$,
\item
\ $\forall r_j^l[x] \in \bigcup_{k = 1}^n T_k: \exists i \in \{ 1, \ldots, n \}: h(r_j^l[x]) = r_j^l[x_i]$,
\item
\ $\forall i \in \{ 1, \ldots, n \}: \forall p, q \in T_i: p <_i q \Rightarrow h(p) < h(q)$,
\item
\ $\forall r_j^l[x] \in \bigcup_{k = 1}^n T_k: h(r_j^l[x]) = r_j^l[x_i] \Rightarrow
(i = 0 \vee \exists w_i[x_i] \in H: w_i[x_i] < r_j^l[x_i])$,
\item
\ $\forall r_j^l[x] \in \bigcup_{k = 1}^n T_k: (h(r_j^l[x]) = r_j^l[x_i] \wedge i \neq j \wedge c_j \in H) \Rightarrow c_i \in H$.
\end{itemize}
An $x_i$ is called a \emph{version} of the data element $x$.
\end{define}

The above definition assumes that prior to any write operation, there already
exists an initial version $x_0$ for every data element $x$.

Mainly for consistency reasons, multiversion histories maintain the superscripts
of read operations as introduced by Definition \ref{transaction}.
Apart from this, the here defined multiversion histories differ from the ones in \cite{BHG87} because
$h$ is not expected to map transaction operations $r_i^l[x]$ with $w_i[x] <_i r_i[x]$ to $r_i^l[x_i]$.
The criterion would be too restrictive for the definition of serializable MC-histories from Section \ref{interpret}.
However, for serializable multiversion histories, we still accomplish a similar result as in \cite{BHG87} because
the definition of multiversion serializability graphs from below accounts for this issue.

\begin{define}
Let $D$ be the set of data elements of all operations of a multiversion history $H$,
so $D = \{ x\ |\ \exists r_i[x_j] \in H \vee \exists w_i[x_i] \in H \}$.
A \emph{version order} $\ll$ establishes for every data element $x \in D$ a total order of its versions, such
that $x_0$ is the smallest version:
\[\forall x \in D: \forall i,j \in \mathbb{N} \setminus \{ 0 \}: x_0 \ll x_i \wedge (i \neq j \Rightarrow x_i \ll x_j \vee x_j \ll x_i).\]
A version order that adheres to the following predicate is called \emph{write version order}:
\[\forall w_i[x_i], w_j[x_j] \in H: (w_i[x_i] < w_j[x_j] \vee i = 0) \Rightarrow x_i \ll x_j.\]
\end{define}

Write version orders are specific version orders. As we will see, it turns out that
we have to rely on write version orders in order to create a serializability theory for MC-histories.

To keeps things short, we omit the definition of serializable (or more specifically 1-serializable) multiversion histories.
Instead, we turn to the definition of multiversion serizalizability graphs straight away and assume
that the reader is familiar with the underlying serializability theorem (see \cite{BHG87}).

\begin{define}
\label{mvsg}
Let $H$ be a multiversion history for the $rw$-transactions $\{T_1, \ldots, T_n\}$ and $\ll$ be a corresponding version order.
The \emph{multiversion serializability graph} $MVSG \subseteq \{ T_1, \ldots, T_n \}^2$ for $H$ and $\ll$ is
given be the following predicate:
\begin{gather*}
(T_i, T_j) \in MVSG :\Leftrightarrow
c_i \in T_i \wedge c_j \in T_j \wedge \exists r_k^h[x_l], w_m[x_m] \in H:\\
(i = j = k = m \wedge i \neq l \wedge w_i[x_i] < r_i^h[x_l])\ \vee\ (i \neq j \wedge m = i = l \wedge k = j) \vee\\
(i \neq j \wedge m = i \wedge l = j \wedge x_m \ll x_l) \vee\ (i \neq j \wedge k = i \wedge m = j \wedge x_l \ll x_m).
\end{gather*}

Instead of writing $(T_i, T_j) \in MVSG$ we simply write $T_i \rightarrow T_j$.
If one of the last two disjunctive clauses holds, then $T_i \rightarrow T_j$ is called a \emph{version order edge}.
\end{define}

Since Definition \ref{multversionhist} enables multiversion histories with operations $w_i[x_i] < r_i^l[x_j]$ and $i \neq j$, the first
disjunctive clause in Definition \ref{mvsg}
introduces graph edges for just this case. In other words: $w_i[x_i] < r_i^l[x_j], i \neq j$ is impossible
for committing transactions $T_i$ and $T_j$ if $MVSG$ is acyclic.

\subsection{Interpretation of MC-Histories}
\label{interpret}

Intuitively, not all serial MC-histories should be considered serializable.
To understand this, let us reconsider $H_1$ from above:
$m_3^{1,4}$ accesses a cached method result which is based on the version of $x$
such as read by $T_1$. However, in the meantime, $T_2$ wrote $x$ and might
have created a new value for it. Further, $r_3^5[x]$ read the value of $x$ written by $T_2$.
This means that $m_3^{1,4}$ refers to another value of $x$ than $r_3^5[x]$, although
this should not be the case. Still $H_1$ is serial.
If the method call that caused $m_3^{1,4}$ had not been a cache hit but had been executed
normally, it would have read $x$ by some operation $r_3^k[x]$. And this value would have
been the value written by $T_2$.

The conventional definition for serializable 1-version histories is based on
the serializability of serial histories. Unfortunately as just seen, this approach is
not applicable to MC-histories.
Then what is a good definition of serializability for MC-histories?
As a solution we will interpret MC-histories as multiversion histories
by means of an embedding function $MV$.
$MV$ maps all operations of an MC-history
to one or more multiversion operations. This way $MV$ produces a multiversion history
that \emph{exactly} reflects all the conflicts that exist for $H$.

Let us begin with an example to convey these intentions. Assume $H_1$ from above
is mapped to the following multiversion history:
\[MV(H_1) = r_1^4 [y_0] r_1^4[x_0] c_1 w_2[x_2] c_2 r_3^1[y_0] r_3^1[x_0] r_3^5[x_2] c_3.\]
The original operations $r_1^4[x] r_1^4[y]$ are mapped to $r_1^4 [y_0] r_1^4[x_0]$ where
$y_0$ and $x_0$ state the versions that these operations read.
$m_3^{1,4}$ is mapped to $r_3^1 [y_0] r_3^1[x_0]$ since it essentially accesses the same
versions of $x$ and $y$ as the read operations to which it refers in $H_1$
(namely $r_1^4[x]$ and $r_1^4[y]$). The superscript for $r_3^1 [y_0]$ and $r_3^1[x_0]$
has been chosen more or less arbitrarily -- because of $r_3^5[x_2]$, it must not equal $5$.
(The superscript is only required for conformance with Definition \ref{multversionhist}.)
Finally $w_2[x_2]$ just writes a respective new version of $x$ and relates to
$w_2[x]$ from $H_1$.

In the following, we will generalize the interpretation function $MV$. Thus we can define an MC-history $H$ to be
serializable if and only if $MV(H)$'s multiversion serialization graph is acyclic for a write version order.
E.g. $MV(H_1)$'s multiversion serializability graph is cyclic for the version order $x_0 \ll x_2$.
It contains the version order edges $T_1 \rightarrow T_2$ (due to $r_1^4[x_0]$ and $w_2[x_2]$),
$T_3 \rightarrow T_2$ (due to $r_3^1[x_0]$ and $w_2[x_2]$) as well as the edge $T_2 \rightarrow T_3$
(due to $w_2[x_2]$ and $r_3^5[x_2]$).
This suits our intuition not to consider $H_1$ as serializable.

For $MV$ it is crucial that it maps \emph{all conflicts} of
an MC-history $H$ to $H$'s multiversion image. Otherwise one might obtain a
multiversion history $MV(H)$ that is 1-serializable although its origin $H$ should
not be considered serializable. The resulting formalism for MC-histories would then
lead to serialization protocols that do not create truly serializable histories.
E.g. the history
\[H_3 = r_1^4[y]r_1^4[x]c_1 w_2[x]c_2 m_3^{1,4}w_3[x]c_3\]
should not be considered
serializable for similar reasons as $H_1$. However, a naive mapping of $H_3$ like
\[r_1^4 [y_0] r_1^4[x_0] c_1 w_2[x_2] c_2 r_3^1[y_0] r_3^1[x_0] w_3[x_3] c_3\]
is 1-serializable but ignores the conflict $w_2[x] \nparallel w_3[x]$ in $H_3$ because
the respective operations $w_2[x_2]$ and $w_3[x_3]$ do not conflict.
So $MV$ has to be defined in way such that this conflict is reflected in $MV(H_3)$.
An appropriate definition of $MV$ results in: $MV(H_3) =$
\[ r_1^4 [y_0] r_1^4[x_0] c_1 w_2[x_2] c_2 r_3^1[y_0] r_3^1[x_0] \begin{array}{ll} \nearrow r_3^2[x_2] \searrow \\
                                              \searrow w_3[x_3] \nearrow \end{array} c_3.\]
Here, the operation $r_3^2[x_2]$ has been introduced to ensure that the set of conflicts
in respect to transactions from $H_3$ and $MV(H_3)$ remain identical.
The next definition states the general structure of $MV$.

\begin{define}
\label{mv}
Let $H$ be an MC-history with the transactions $\mathbb{T} = \{ T_1, \ldots, T_n \}$.
The function
\[V: H \rightarrow \{1, \ldots, n \}, V(p) \mapsto k\]
shall return the index $k$
of the last write operation $w_k[x] \in H$ before $p$ such that $c_k \in H$. If no such $w_k[x]$ exists,
$V(p)$ shall be zero, so $V(p) = 0$.
Further, the function
\[ss : \mathbb{N} \times \mathbb{N} \times \mathbb{T} \rightarrow \mathbb{N}, (i, j, T_i) \mapsto h\]
shall return a unique number for an argument $(i, j, T_i)$ such that $h \notin \{ k | r_i^k[x] \in T_i \}$.\footnote{
The specific structure of $ss$ is not of interest. Below, it is just required
to produce unique superscripts for read operations in respect to a transaction $T_i$.}

The \emph{interpretation function} $MV$ is then defined be means of an auxiliary function $mv$ with\vspace{0.1cm}
\\
$mv(r_i^k[x]) = \{ r_i^k[x_{V(r_i^k[x])}] \}$,\ $mv(w_i[x]) = \left\{\begin{array}{l} \{ w_i[x_i] \}  \textrm{\ if\ \ } \exists r_i^k[x] \in H: r_i^k[x] < w_i[x]\vspace{0.1cm}\\
 \{ w_i[x_i], r_i^k[x_{V(w_i[x])}] \}  \textrm{\ otherwise,} \end{array} \right .$\vspace{0.1cm}\\
$mv(m_i^{k,j}) = \{ r_i^h[x_{V(q)}]|q = r_k^j[x] \in H \wedge h = ss(k,j,T_i) \}$ and $MV(H)= \cup_{p \in H} mv(p)$.\vspace{0.1cm}

The partial ordering relation $<'$ for $MV(H)$ is inherited from $H$'s partial ordering relation $<$,
more specifically: $p <' q :\Leftrightarrow$
\[\big(mv^{-1}(p) < mv^{-1}(q) \vee (\{p,q\} \subseteq mv(m_i^{j,k})\ \wedge\
 p = r_j^k[x_s] \wedge q = r_j^k[y_t] \wedge r_j^k[x] < r_j^k[y])\big).\]
\end{define}

The latter part of the definition of $<'$ deals with ordering read operations that replace method operations.
$MV$ produces a well formed multiversion history according to Definition \ref{multversionhist}.
The next theorem shows that for an $rw$-history $H$, $MV$ produces a multiversion history with
(practically) the same serialization graph as $H$.

\begin{theorem}
\label{sgmvsg}
Let $H$ be an $rw$-history.
Further, $SG^*(H)$ shall be the transitive closure of the 1-version serializability graph of $H$ (according to \cite{BHG87})
and $MVSG^*(MV(H))$ shall be the transitive
closure of the multiversion
serializability graph of $MV(H)$ with some write version order.
Then, the two graphs are identical, so $SG^*(H) = MVSG^*(MV(H))$.
\end{theorem}

\begin{proof}
Obviously, $\forall i \in \{ 1, \ldots, n \}: c_i \in T_i \Leftrightarrow c_i \in mv(T_i)$ holds.
This means that conditions for graph edges that request participating transactions
to be committed do not have to be considered any further for this proof.

"$\subseteq$": Let $T_i \rightarrow T_j$ be in $SG$.
Then, there are operations $p \in T_i$, $q \in T_j$ with $p < q$, $p \nparallel q$ and $i \neq j$.
Moreover, there is an $x$ with $\{ x \} = d(p) \cap d(q)$.

If $a(p) = r, a(q) = w$ one has got $r_i^h[x_s] <' w_j[x_j]$ in $MV(H)$ (for some $s$).
Thus, $w_s[x] < w_j[x]$ must hold and so $x_s \ll x_j$. This leads to the version order
edge $T_i \rightarrow T_j \in MVSG$.
If $a(p) = w, a(q) = r$, one has got $w_i[x_i] <' r_j^h[x_s]$ in $MV(H)$ (for some $s$) with the following
two options for $w_s[x_s]$:
Either one obtains the trivial case $i = s$ or $w_i[x] < w_s[x]$. $w_s[x] < w_i[x]$ cannot hold
because it would lead to $w_i[x_i] <' w_s[x_s]$ and so $r_j^h[x_i]$ because
in Definition \ref{mv} the index $i$ is determined by $V$ (contradiction).
Since $c_s \in H$ (according to the definition of $V$), $T_s \rightarrow T_j$ is in $MVSG$.
As one will see as part of the next case, $w_i[x] < w_s[x]$ implies the edge $T_i \rightarrow T_s \in MVSG^*$
and so $T_i \rightarrow T_j \in MVSG^*$.

Finally, consider $a(p) = w, a(q) = w$:
Let $w_i[x] = w_{k_1}[x] < \ldots < w_{k_n}[x] = w_j[x]$ be the sequence of \emph{all}
write operations in $H$ in respect to $x$ between $w_i[x]$ and $w_j[x]$ such that $n \geq 2$ and
$c_{k_o} \in T_o$ for all $o \in \{ 1, \ldots, n \}$.
Next we prove that there is a path $T_{k_1} \rightarrow T_{k_n} \in MVSG^*$ by induction on $n$.
$n = 2$: For this case $mv(w_{k_2}[x]) = \{ w_{k_2}[x_{k_2}], r_{k_2}[x_{k_1}] \}$ due to the definition of $V$ and also
$w_{k_1}[x_{k_1}] <' r_{k_2}[x_{k_1}]$. Thus, $T_{k_1} \rightarrow T_{k_2} \in MVSG$.
$n - 1 \curvearrowright n$: The argument is analogous to the case $n = 2$.
The only difference is to replace $k_1$ by $k_{n-1}$ and $k_2$ by $k_n$.

"$\supseteq$":
Let $T_i \rightarrow T_j$ be in $MVSG$. $T_i \rightarrow T_j$ can be a version order edge or
an edge due to $w_i[x_i] <' r_j^h[x_i]$ with $i \neq j$.
In particular the case $w_i[x_i] <' r_i^h[x_l]$ with $i \neq l$ (from the first disjunctive clause
of Definition \ref{mvsg}) can be excluded because of $mv$'s Definition.

Consider the case $w_i[x_i] <' r_j^h[x_i]$:
According to the definition of $mv$ one has got $w_i[x] < r_j^h[x]$ (if $mv(r_j[x]) = \{ r_j^h[x_k] \}$ for some $k$)
or $w_i[x] < w_j[x]$ (if $mv(w_j[x]) = \{ r_j^h[x_k], w_j[x_j] \}$ for some $k$). So $T_i \rightarrow T_j \in SG$.
$r_j^h[x_i]$ cannot be in the range of a method operation because $H$ is an $rw$-history.

If $T_i \rightarrow T_j \in MVSG$ is a version order edge one has got two cases.
Case 1: $w_i[x_i], r_k^h[x_j] \in MV(H)$ (for some $k$) with $x_i \ll x_j$.
$i \neq 0$ holds because of $w_i[x_i]$ and because $\ll$ is a version order. Thus, $j > 0$, which implies that
a $w_j[x_j]$ exists in $MV(H)$. Since $\ll$ is a write version order, $w_i[x] < w_j[x]$ follows and further,
$T_i \rightarrow T_j \in SG$ follows.

Case 2: One has got two operations $r_i^h[x_k], w_j[x_j] \in MV(H)$ (for some $k$) with $x_k \ll x_j$.
There two are subordinate cases, namely $r_i^h[x_k] <' w_j[x_j]$ and $w_j[x_j] <' r_i^h[x_k]$.
(The two operations can be compared by means of $<'$,
since their preimages $p,q \in H$ in respect to $mv$ must be conflicting and so $p < q$ or $q < p$,
but this relationship is maintained by $<'$.)
Consider $r_i^h[x_k] < w_j[x_j]$ first. Then, $r_i^h[x_k] \in mv(r_i^h[x])$ or $r_i^h[x_k] \in mv(w_i[x])$ and
$r_i^h[x] < w_j[x]$ respectively $w_i[x] < w_j[x]$ follows. So $T_i \rightarrow T_j \in SG$.
($mv^{-1}(r_i^h[x_k]$) cannot be a method operation because $H$ is an $rw$-history.)
Secondly, consider $w_j[x_j] <' r_i[x_k]$. Due to the definition of $V$, $k$ cannot be zero and so,
with $x_k \ll x_j$ one obtains $w_k[x] < w_j[x]$. If $r_i[x_k] \in mv(r_i[x])$ holds, it follows that
$w_k[x] < w_j[x] < r_i[x]$ which implies $V(r_i^h[x]) \neq k$.
This is a contradiction to $r_i[x_k] \in mv(r_i[x])$.
Finally, if $r_i[x_k] \in mv(w_i[x])$ one obtains
$w_k[x] < w_j[x] < w_i[x]$ and thus $V(w_i[x]) \neq k$. However, this also contradicts $r_i[x_k] \in mv(w_i[x])$.
The previous considerations have covered all cases for edges $T_i \rightarrow T_j \in MVSG$.
\end{proof}

Using Definition \ref{mv} one can interpret MC-histories as ordinary multi-version histories.
However, an MC-history does not exhibit the same complexity as its underlying multi-version history.
(E.g. MC-histories without $m$-operations may be considered as ordinary one version-histories.)
Therefore the introduction of $m$-operations greatly simplifies the development $m$-scheduler protocols.

Theorem \ref{sgmvsg} stated that the chosen interpretation function $MV$ is appropriate when applied to an $rw$-history $H$,
since $MV(H)$ essentially holds the same serializability graph as $H$.
Moreover, $MV$ interprets an $m$-operation as a set of read operations accessing just the versions
of data elements which were used when the respective cached method result was first computed.
These facts justify the following definition of serializable MC-histories.

\begin{define}
\label{mcserializable}
An MC-history $H$ is \emph{MC-serializable} iff $MVSG(MV(H))$ is acyclic in respect to some write version order.
\end{define}

E.g. $H_1$ and $H_3$ from above are not MC-serializable because the corresponding multiversion serializabilty graph is cyclic
(and $x_0 \ll x_2$ matches the write version order predicate).

\subsection{Serializability Theorem for MC-Histories}

Using Definition \ref{mcserializable} one can decide whether
an MC-history $H$ is MC-serializable by computing $MV(H)$ and then
checking the resulting history's multiversion serializability graph for cycles.
Clearly, it would be more convenient if we had
a serializability theorem which applies right to $H$ instead of $MV(H)$. The next
definition states how a respective graph should be constructed for $H$.

\begin{define}
\label{defmcsg}
Let $H$ be an MC-history for the transactions $\{T_1, \ldots, T_n\}$.
The \emph{MC-serializability graph} $MCSG \subseteq \{ T_1, \ldots, T_n \}^2$ for $H$ is
given by the following predicate:
\begin{gather*}
(T_i, T_j) \in MCSG:\Leftrightarrow c_i \in T_i \wedge c_j \in T_j\ \wedge\\
\Big( (\exists p \in T_i: \exists q \in T_j: a(p) \neq m\ \wedge\ a(q) \neq m \wedge p \nparallel q \wedge p < q)\ \vee\\
\big(\exists m_i^{k,l},w_j[x],r_k^l[x] \in H: r_k^l[x] < w_j[x]\ \wedge\ (i \neq j \vee w_j[x] < m_i^{k,l})\big)\ \vee\\
(i \neq j \wedge \exists w_i[x],m_j^{k,l},r_k^l[x] \in H^{MC}: w_i[x] < r_k^l[x]) \Big).
\end{gather*}
Instead of $(T_i, T_j) \in MCSG$ we simply write $T_i \rightarrow T_j$.
\end{define}

Consider $H_1$ from above.  Its MC-serializability graph consist of $T_1 \rightarrow T_2$ (due to $r_1^4[x] \nparallel w_2[x]$),
$T_2 \rightarrow T_3$ (due to $w_2[x] \nparallel r_3^5[x]$) and $T_3 \rightarrow T_2$ (due to $w_2[x] \nparallel m_3^{1,4}$).
These are the same edges as in $MVSG(MV(H_1))$ (with $x_0 \ll x_2$).
This observation gives rise to proving the serializability theorem for MC-histories which is stated next.

\begin{theorem}
\label{mcsertheorem}
Let $H$ be an MC-history.
$MCSG^*(H)$ shall be the transitive closure of its MC-serializability graph of $H$
and $MVSG^*(MV(H))$ shall be the transitive
closure of the multiversion serializability graph of $MVSG(H)$ in respect to some write version order.
Then, the two graphs are identical, so $MCSG^*(H) = MVSG^*(MV(H))$.
\end{theorem}

\begin{proof}
Just as for the proof of Theorem \ref{sgmvsg}, conditions for graph edges that request participating transactions
to be committed do not have to be considered any further.

"$\subseteq$": Let $T_i \rightarrow T_j$ be in $MCSG$. Due to the first disjunctive clause of Definition \ref{defmcsg}
$SG(RW(H)) \subseteq MCSG(H)$ holds.
(Just compare the first disjunctive clause of Definition \ref{defmcsg} with the definition of SG from \cite{BHG87}.)
So, if $T_i \rightarrow T_j \in SG(RW(H))$ then $T_i \rightarrow T_j \in MVSG^*(MV(RW(H))) \subseteq MVSG^*(MV(H))$.
(This follows from Theorem \ref{sgmvsg}.)

Now, let $T_i \rightarrow T_j$ be in $MCSG(H) \setminus SG(RW(H))$.
$T_i \rightarrow T_j$ can only exist because of the second or the third disjunctive clause of Definition \ref{defmcsg}.
This means that there are either operations $m_i^{k,l}$, $w_j[x]$, $r_k^l[x]$ with $r_k^l[x] < w_j[x]$ or
operations $w_i[x]$, $m_j^{k,l}$, $r_k^l[x]$ with $w_i[x] < r_k^l[x]$.

For the first case, consider the image in respect to $mv$:
$mv(r_k^l[x]) = \{ r_k^l[x_s] \}$, $w_j[x_j] \in mv'(w_j[x])$
and $r_i^h[x_s] \in mv'(m_i^{k,l})$ (for some $s$). With $r_k^l[x_s] <' w_j[x_j]$ it turns out that
$s = 0 \vee w_s[x_s] <' w_j[x_j]$ and one gets the version order $x_s \ll x_j$.
For this case $i \neq j$ and the operations $r_i^h[x_s]$ and $w_j[x_j]$ result in the version order edge
$T_i \rightarrow T_j \in MVSG$ (see last disjunctive clause of Definition \ref{mvsg}).
If otherwise $i = j$ holds, it follows that $w_j[x_j] = w_i[x_i] <' r_i[x_s]$ for the second
disjunctive clause of Definition \ref{defmcsg}.
Since $i \neq s$, one obtains $T_i \rightarrow T_j = T_i \in MVSG$ because of the first disjunctive clause
of Definition \ref{mvsg}.

If there are operations $w_i[x]$, $m_j^{k,l}$, $r_k^l[x]$ with $w_i[x] < r_k^l[x]$
that cause $T_i \rightarrow T_j \in MCSG$, then their images in respect to $mv$ behave as follows:
$w_i[x_i] <' r_k[x_s] <' r_j^h[x_s]$ (for some $s$). The case $i = s$ is trivial.
Otherwise one can conclude by induction as in the proof of Theorem \ref{sgmvsg} that
$T_i \rightarrow T_s \in MVSG^*$ with $w_s[x_s] \in T_s$.
Thus $T_i \rightarrow T_j$ is in $MVSG^*$. (Note that $s > 0$ because of $V$'s definition and because of
$w_i[x_i]$.)

"$\supseteq$":
Let $T_i \rightarrow T_j$ be in $MVSG(MV(H))$.
Theorem \ref{sgmvsg} has already considered all edges that relate to
conflicts between read and write operations but not method operations.
Therefore, it suffices to analyze edges in $MVSG$ that are cause by the additional images of method operations in respect to $mv$.
So, let $r_n^h[x_s] \in mv'(m_n^{k,l})$ and $w_t[x_t]$ be operations that causes a respective edge $T_i \rightarrow T_j \in MVSG$.
According to Definition \ref{mvsg} one has to distinguish for cases:
$i = j = n = t, i \neq s, w_i[x_i] <' r_i[x_s]$ or
$i \neq j, i = s = t, j = n$ or $i \neq j, i = t, j = s, x_t \ll x_s$
or $i \neq j, n = i, t = j, x_s \ll x_t$.

In the first case, one has got operations $r_k^l[x] < w_i[x] < m_i^{k,l}$ or $w_i[x] < w_o[x] < r_k^l[x] < m_i^{k,l}$, since
otherwise $i = s$ would hold.
$r_k^l[x] < w_i[x] < m_i^{k,l}$ results in the edge $T_i \rightarrow T_i \in MCSG$ with $i = j$ from the
second disjunctive clause of Definition \ref{defmcsg}.
$w_i[x] < w_o[x] < r_k^l[x] < m_i^{k,l}$ results in $T_i \rightarrow T_o \rightarrow T_i \in MCSG$.

The second case leads to $w_i[x_i] <' r_j^h[x_i] \in mv(m_j^{k,l})$ with $w_i[x_i] \in mv(w_i[x])$.
Therefore, there exists an $r_k^l[x]$ with $w_i[x] < r_k^l[x]$ in $H$.
If $r_k^l[x] < w_i[x]$ would hold, applying $mv$ would return $mv(r_k^l[x]) = \{ r_k[x_g] \}$ for some $g \neq i$.
This would lead to $r_j[x_g] \in mv(m_j^{k,l})$ instead of $r_j[x_i] \in mv(m_j^{k,l})$ (contradiction).
Thus, $T_i \rightarrow T_j \in MCSG$ follows from the last disjunctive clause of Definition \ref{defmcsg}.

Considering the case $i \neq j, i = t, j = s, x_t \ll x_s$:
Here, $w_i[x] < w_j[x]$ follows right away because $\ll$ is a write version order.
(Note that $t = i$ cannot be zero.)

The last case creates the situation $x_s \ll x_j$, $r_i^h[x_s] \in mv(m_i^{k,l})$
and $w_j[x_j] \in mv(w_j[x])$ with $w_j[x] \in H$.
Moreover, due to $m_i^{k,l}$, there must be a $r_k^l[x] \in H$ with $r_k^l[x] < m_i^{k,l}$.
If $r_k^l[x] < w_j[x]$ holds, one obtains $T_i \rightarrow T_j$ for the second disjunctive clause of Definition \ref{defmcsg}.
Now consider $w_j[x] < r_k^l[x]$: If $r_k^l[x]$ reads from $T_j$, applying $mv$ results in
$w_j[x_j] <' r_k^l[x_j] <' r_i^h[x_j] \in mv(m_i^{k,l})$ and so $j = s$ but this is a contraction to $x_s \ll x_j$.
Otherwise $r_k^l[x]$ reads $x$ from a $T_o \neq T_j$ and one has got $w_j[x] < w_o[x] < r_k^l[x]$.
Applying $mv$ results in $w_j[x_j] <' w_o[x_o] <' r_k^l[x_o] <' r_i^h[x_o] \in mv(m_i^{k,l})$.
Thus, $s = o$ and finally $x_j \ll x_s$ follows (because of $w_j[x_j] < w_o[x_o]$). However, this contradicts the case's precondition.
\end{proof}

Given an MC-history $H$, Theorem \ref{mcsertheorem} confirms that the transitive closure of $H$'s MC-serializability graph
is identical to the transitive closure of $MV(H)$'s multiversion serializability graph.
Since a transitive closure does neither add nor remove
graph cycles, we can indeed rely on Definition \ref{defmcsg} to check for MC-serializability.

\section{Recovery for MC-Histories}
\label{recoverysec}

Before developing a serializability protocol for transactional method caching, we want
to address the simpler task of creating a recovery protocol. In this respect, we are
interested in applying conventional recovery qualities such as "recoverable" or "strict".
Again, the definition of these qualities must be adapted to the structure of MC-histories.
This section defines the corresponding qualities and gives a lemma on which an $m$-scheduler's
recovery protocol can be based. The second part of this section discusses the protocol's
implementation.

\subsection{Formalism}

\begin{define}
\label{reads}
Let $H$ be an MC-history with the transactions $\mathbb{T} = \{ T_1, \ldots, T_n \}$.
A transaction $T_i \in \mathbb{T}$ \emph{reads (a data element) $x$ from $T_j \in \mathbb{T}$ via an operation $p \in T_j$} iff:
\begin{gather*}
\exists r_h^k[x],w_j[x] \in H: w_j[x] < r_h^k[x]\ \wedge\ (h = i \vee m_i^{h,k} \in H) \wedge  \neg (a_j < r_h^k[x])\ \wedge\\
\forall w_o[x] \in H: w_j[x] < w_o[x] < r_h^k[x] \Rightarrow a_o < r_h^k[x].
\end{gather*}
We have $p = r_h^k[x]$, if $h = i$ holds for the given predicate and $p = m_i^{h,k}$ otherwise.
The relationship between $T_i$, $x$, $T_j$ and $p$ is expressed by $reads(T_i, x, T_j, p)$.
$reads$ forms the so called \emph{reads-from-relation}.
\end{define}

For the MC-history
\[H_4 = w_2[x] r_1^1[y] r_1^1[x] c_1  c_2 m_3^{1,1} w_3[x] c_3\] we have
$reads = \{ (T_1, x, T_2, r_1^1[x]), (T_3, x, T_2, m_3^{1,1}) \}$. Using the reads-from-relation,
most conventional recovery qualities can also be applied to MC-histories.

\begin{define}
\label{recovery}
An MC-history $H$ with the transactions $\mathbb{T} = \{ T_1, \ldots, T_n \}$
and the data elements $D$
is \emph{recoverable} respectively \emph{ACA (avoiding cascading aborts)} respectively \emph{strict}, iff
the following qualities hold:
\begin{itemize}
\item\ \emph{recoverable}: \[\forall i,j \in \{1,\ldots,n\}: \forall x \in D: \forall p \in H: \big(i \neq j \wedge
reads(T_i, x, T_j, p) \wedge c_i \in H\big) \Rightarrow c_j < c_i,\]

\item\ \emph{ACA}:
\[\forall i,j \in \{1,\ldots,n\}: \forall x \in D: \forall p \in H: \big(i \neq j \wedge reads(T_i, x, T_j, p)\big) \Rightarrow c_j < p,\]

\item\ \emph{strict}: $H$ is ACA and
\[\forall w_i[x], w_j[x] \in H: (i \neq j \wedge w_j[x] < w_i[x]) \Rightarrow (a_j < w_i[x] \vee c_j < w_i[x]).\]
\end{itemize}
\end{define}

Obviously, the standard inclusion statement "strict $\subset$ ACA $\subset$ recoverable"
also is true for MC-histories.
The four MC-histories $H_5$ to $H_8$, which are presented next, only differ in respect to
the placement of $c_1$ but:
$H_5$ is not recoverable, $H_6$ is recoverable but not ACA,
$H_7$ is ACA but not strict, $H_8$ is strict.
\begin{gather*}
H_5 = w_1[x] w_1[y] w_2[y] r_2^1[x] m_3^{2,1} c_3 c_1 c_2,\\
H_6 = w_1[x] w_1[y] w_2[y] r_2^1[x] m_3^{2,1}c_1 c_3 c_2,\\
H_7 = w_1[x] w_1[y] w_2[y] c_1 r_2^1[x] m_3^{2,1} c_3 c_2,\\
H_8 = w_1[x] w_1[y] c_1 w_2[y] r_2^1[x] m_3^{2,1} c_3 c_2.
\end{gather*}

% Man betrachte außerdem die folgende Methoden-Cache-Historie $(H_5, <)$:
% \[ (H_5, <) = r_1^1[x] r_2[x] w_3[x] m_2^{1,1} c_3 c_2 c_1.\]
% $(H_5, <)$ ist strikt, obwohl zwischen $w_3[x]$ und $m_2^{1,1}$ kein $c_3$ liegt!
% Hier ist \emph{ausschließlich} die Liest-Von-Relation aus der Kaskadenfreiheit ausschlaggebend,
% und in dieser kommt $reads(T_2, x, T_3, m_2^{1,1})$ nicht vor, weil $r_1^1[x]$ \emph{nicht} von
% $T_3$ liest.
The next lemma states how an $m$-scheduler can ensure that together with the $rw$-scheduler, it
produces ACA MC-histories. By requesting an MC-history's $rw$-projection to be ACA the lemma
assumes that the $rw$-scheduler will already provide ACA \emph{$rw$-histories}. Given that the
$m$-scheduler guarantees an additional predicate, the joint MC-history will be ACA too.

\begin{lemma}
\label{rwkaskadenfrei}
Let $H$ be an MC-history for the transactions $\mathbb{T} = \{T_1, \ldots, T_n\}$ and let the following
predicate hold:
\[\forall T_i \in \mathbb{T}: \forall x \in D: reads(T_i, x, T_i, r_i^l[x]) \Rightarrow
\forall m_j^{i,l} \in H: i \neq j \Rightarrow c_i < m_j^{i,l}.\]
Then, $H$ is ACA iff $RW(H)$ is ACA.
\end{lemma}

\begin{proof}
"$\Rightarrow$": Let $H$ be ACA. Since $RW(H) \subseteq H$ holds, the reads-from-relation of $RW(H)$
is a subset of $H$'s reads-from-relation. Therefore $RW(H)$ is also ACA.

"$\Leftarrow$": Let $RW(H)$ be ACA.
Thus, in respect to $H$ only the additional method operations might violate ACA.
Let $m_j^{k,l} \in H$ be such a method operation that reads from $T_i$ via $w_i[x]$,
so $reads(T_j, x, T_i, m_j^{k,l})$ holds. Due to Definition \ref{reads} there must also
be an $r_k^l[x]$ with $reads(T_k, x, T_i, r_k^l[x])$. Further, $r_k^l[x] < m_j^{k,l}$ must hold
because of Definition \ref{mchist}.
If $k \neq i$ then $c_i < r_k^l[x] < m_j^{k,l}$ follows, since $RW(H)$ is ACA.
Otherwise, one obtains $reads(T_i, x, T_i, r_i^l[x])$ and so $c_i < m_j^{k,l}$ if $i \neq j$
due to the Lemma's predicate. In either case $H$ is ACA.
\end{proof}

The next MC-history shows that the predicate of Lemma \ref{rwkaskadenfrei} is necessary:
\[H_9 = w_1[x]r_1^1[x]m_2^{1,1}c_1c_2\]
is not ACA because of $reads(T_2,x,T_1,m_2^{1,1})$ and $m_2^{1,1} < c_1$.
However, $RW(H_9) = w_1[x]r_1^1[x]\\
c_1c_2$ is ACA.

As the following example shows, Lemma \ref{rwkaskadenfrei} cannot be rephrased for MC-histories that are just
recoverable:
\[H_{10} = w_1[x] r_2^1[x] m_3^{2,1} c_3 c_1 c_2\]
is not recoverable, since
the relation $reads(T_3, x, T_1, m_3^{2,1})$ holds and $c_3 < c_1$.
Still, $RW(H_{10})$ is recoverable.

If one wants MC-histories to be strict and not just ACA,
it suffices to keep the predicate from Lemma \ref{rwkaskadenfrei} and to expect
the $rw$-scheduler to produce strict $rw$-histories:

\begin{lemma}
\label{rwstrikt}
An MC-history $H$ is strict iff $H$ is ACA and $RW(H)$ is strict.
\end{lemma}
%The proof for this Lemma is straight forward and follows from Lemma \ref{rwkaskadenfrei}.

\begin{proof}
"$\Rightarrow$": Let $H$ be strict. Since $RW$ does not remove any commit or abort operations
$RW(H)$ must be strict too.

"$\Leftarrow$": Let $H$ be ACA and $RW(H)$ be strict. In respect to $H$ only method operations
must be checked. However, additional method operations do not impact the strictness predicate for write operations
from Definition \ref{recovery}.
\end{proof}

\subsection{Implementation}

We now describe a simple protocol that produces ACA respectively strict $MC$-histories given
that the $rw$-scheduler creates ACA respectively strict $rw$-histories.
As stated by Lemma \ref{rwstrikt}
and \ref{rwkaskadenfrei} the $m$-scheduler's job is just to guarantee the predicate
of Lemma \ref{rwkaskadenfrei}. Surprisingly, this can be done entirely on the client side of
a related system: For every transaction $T_i$ started at the client,
the method cache keeps a flag which indicates whether or not
there has already occurred a write method call inside $T_i$. (For a new transaction the
flag is false, meaning no write method call has occurred yet.)
After the first write method call of $T_i$, every new method call result $r$ which is
computed inside $T_i$ and stored in the method cache, remains locked client
until $T_i$ ends. The lock prevents concurrent transactions from producing a cache hit on $r$
before $T_i$ ends. At $T_i$'s commit, the lock is removed and other transactions may access $r$.
However, if $T_i$ aborts, then $r$ is entirely removed from the cache.

The protocol is correct because $reads(T_i, x, T_i, r_i^l[x])$ from the predicate of
Lemma \ref{rwkaskadenfrei} can only hold, if some write operation has ever occurred in $T_i$.
When this happens, the lock on new cached method results produced by $T_i$ prevents
other transactions from reading those cached method results before $T_i$ has committed.

% \section{Lock Protocol}
%
% \subsection{Formalism}
%
% Example: current
%
% Definition: current
%
% Example: m-locking
%
% Definition: m-locking
%
% Lemma: If RW(H) is produced by a 2PL and H ist aca, m-locking and current then
% H is also MC-serializable.
%
% Proof sketch:
%
% \subsection{Implementation}
%
% Algorithm:
%
% Correctness:

\section{Optimistic Caching Timestamp Protocol}
\label{protocol}

\subsection{Formalism}
\label{protocolformalism}

This section presents an optimstic caching timestamp protocol (OCTP) for scheduling method operations as part of $MC$-histories.
An $m$-scheduler that applies this protocol can be integrated with an $rw$-scheduler that follows a timestamp
protocol itself but also with a strict two-phase lock protocol. An integration with a strict two-phase lock
protocol is possible by interpreting the $rw$-scheduler's commit order as a timestamp order.
(\cite{BHG87} showed that this is legitimate.)

Apart from the protocol presented next, we have developed another serialization protocol
for an $m$-scheduler whose essential idea is related to the one of OCC from \cite{223787}.
For a more compact contribution we do not present this protocol. We prefer to present OCTP mainly because
it is a strong improvement over the OCC-like protocol: It accepts a superset
of histories that the OCC-like protocol accepts\footnote{This can be proven.}
and it causes much lower transaction abortion rates. The latter statement is substantiated
by the experiments from Section \ref{evaluation}.\footnote{The effect can also by explained analytically
but this is beyond the scope of this paper.} As opposed to the OCC-like protocol, the correctness
of OCTP is not straight forward to see.  We will have to make good use of the formalism from Section \ref{theory}
to prove it correct.

The fundamental concept of timestamp protocols are timestamps.
For clearity and completeness we define them next.

\begin{define}
Let $H$ be an MC-history with the transactions $\mathbb{T} = \{ T_1, \ldots, T_n \}$.
$ts : \{ T_1, \ldots, T_n \} \rightarrow \mathbb{N}$ is a \emph{timestamp function} iff
\[\forall i,j \in \{ 1, \ldots, n \}: ts(T_i) = ts(T_j) \Rightarrow i = j.\]
\end{define}

For conventional timestamp protocols conflicting operations should be ordered along the timestamp order of the
transactions to which they belong.

\begin{define}
\label{zgeordnet}
Let $H$ be an MC-history with the transactions $\mathbb{T} = \{ T_1, \ldots, T_n \}$.
$H$ is \emph{$t$-ordered in respect to a timestamp function $ts$} iff
\begin{gather*}
\forall p, q \in H: \forall i,j \in \{ 1, \ldots, n \}:
\big(p \in T_i \wedge q \in T_j \wedge p \nparallel q \wedge ts(T_i) < ts(T_j)\big) \Rightarrow\\
(a_i \in H \vee a_j \in H \vee p < q).
\end{gather*}
\end{define}

It is well known and easy to prove that $t$-ordered $rw$-histories are serializable.
The reason for this is that conflicting read and write operations
dictate the direction of edges in a respective serializability graph.
However, for a method operation that conflicts with a write operation
the direction of a respective edge in the MC-serializability graph does not necessarily depend on the two operation's
order. E.g. $H_1$ from above is $t$-ordered for the timestamp function $ts(T_i) = i$ but the operations
$r_1^4[x] < w_2[x] < m_3^{1,4}$ produce an edge $T_3 \rightarrow T_2$.
Therefore the timestamp rule does not guarantee MC-serializability.

In the following, an edge $T_j \rightarrow T_i$ is called a \emph{reverse edge}, if and only if it
is produced by two conflicting operations $p \in T_i$ and $q \in T_j$ with $ts(T_i) < ts(T_j)$ .
Otherwise we call it a \emph{normal edge}.

Interestingly, if an MC-history $H$ is $t$-ordered, $H$'s reverse edges can only be created by the condition
$\exists m_i^{k,l},w_j[x],r_k^l[x] \in H: (r_k^l[x] < w_j[x] \wedge (i \neq j \vee w_j[x] < m_i^{k,l}))$
from Definition \ref{defmcsg}. This implies that
the read operation $r_k^l[x]$ to which $m_i^{k,l}$ refers must have occurred before $w_j[x]$.

One way to develop a timestamp protocol for MC-histories would be to entirely forbid reverse edges.\footnote{
This approach leads to the OCC-like protocol mentioned at the beginning of this section.}
But we can go a more general way and trade off reverse edges against normal graph edges!
To illustrate this idea, consider the following prefix of $H_1$:
\[r_1^4[y]r_1^4[x]c_1 w_2[x]c_2 m_3^{1,4}.\]
When $m_3^{1,4}$ is scheduled it produces the reverse edge $T_3 \rightarrow T_2$. So afterwards the
scheduler's duty should be to avoid edges from $T_2$ to $T_3$.
At the point of time when the $m$-scheduler accepts $m_3^{1,4}$, $T_3$ has still a (good) chance to commit.
However if we forbade reverse edges entirely, the $m$-scheduler would have to reject $m_3^{1,4}$ and thus abort $T_3$ right away.

\begin{figure}[!tb]
\centering
\includegraphics[width=4cm]{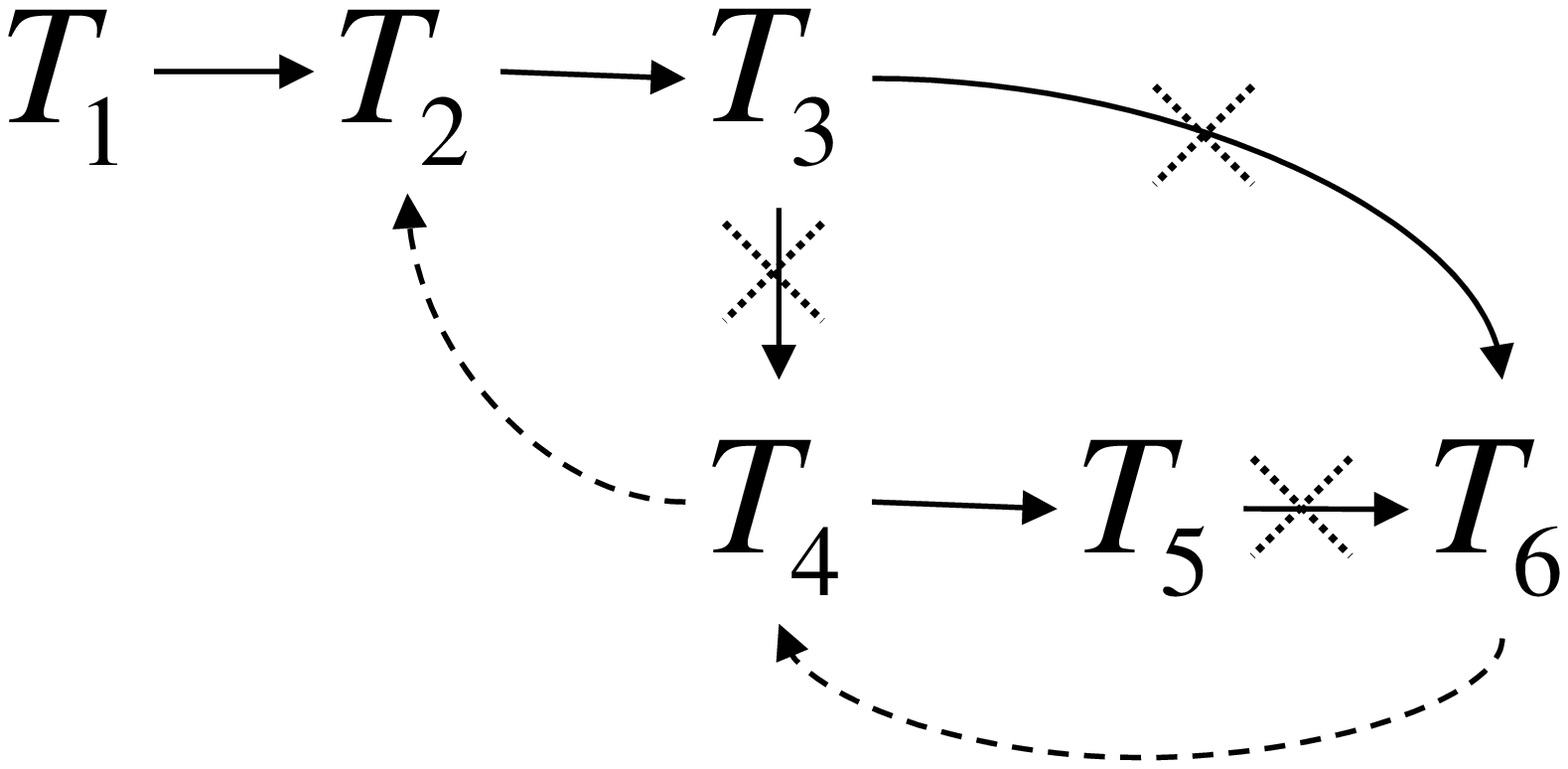}
\caption{An MC-Serializability Graph to Illustrate the Idea behind OCTP}
\label{tsprotokollgraph}
\end{figure}

In general, the following rule should hold:
If the $m$-scheduler accepts a method operation producing a reverse edge $T_i \rightarrow T_j$
then it should ensure that there are no edges $T_h \rightarrow T_i$ with $ts(T_j) \leq ts(T_h)$.
As an example, suppose the graph from Figure \ref{tsprotokollgraph}
was an MC-serializability graph with the timestamp function $ts(T_i) = i$.
The dotted arrows then represent reverse edges.
According to the stated rule, the graph edge $T_5 \rightarrow T_6$ must be excluded because of the reverse edge $T_6 \rightarrow T_4$.
Similarly, $T_3 \rightarrow T_4$ contradicts the rule due to the reverse edge $T_2 \rightarrow T_3$.
But how about $T_3 \rightarrow T_6$? It adheres to the stated rule and still leads to a graph cycle.
Apparently, it does not suffice to consider single reverse edges. Instead, one has to consider
\emph{paths of reverse edges}. In Figure \ref{tsprotokollgraph} a path of reverse
edges starting from $T_6$ leads back to $T_2$. Therefore, no transactions with $ts(T_i) \geq ts(T_2)$ should point
to $T_6$.

The function $ts_{fit}(T_i)$ which is defined next, computes
the minimum timestamp of all those transactions that can be reached from transaction $T_i$ via paths consisting exclusively
of reverse edges. The computation is based on the operation order of an underlying MC-history (prefix) and can
be performed dynamically by the $m$-scheduler. The function forms the basis of a respective serializability protocol.

\begin{define}
\label{zpstempel}
Let $H$ be an MC-history with the transactions $\{ T_1, \ldots, T_n \}$ and a timestamp function $ts$.
The \emph{fitting timestamp} function
\[ts_{fit}: \{ T_i\ |\ i \in \{1, \ldots, n \} \wedge c_i \in T_i \} \rightarrow \mathbb{N}\]
is computed as follows:
\begin{gather*}
ts_{fit}(T_i) = \min \big( \{ ts(T_i) \}\ \cup \\
\{\ ts_{fit}(T_j)\ |\ \exists w_j[x], m_i^{k,l}, r_k^l[x] \in H:
r_k^l[x] < w_j[x] \wedge ts(T_j) < ts(T_i) \wedge c_j \in H\ \} \big).
\end{gather*}
\end{define}

\begin{lemma}
$ts_{fit}$ is well defined.
\end{lemma}

\begin{proof}
Consider $ts_{fit}(T_i)$ according to Definition \ref{zpstempel}.
The argument of $\min(\ldots)$ is a non-empty set, since it contains $ts(T_i)$.
Further, every $T_j$ referenced by the set $\{ ts_{fit}(T_j)\ |\ \ldots \}$ from above
has committed (so $c_j \in T_j$) and lies in the domain of $ts_{fit}$.
For every $T_j$ referenced by $\{ ts_{fit}(T_j)\ |\ \ldots \}$ we have $ts(T_j) < ts(T_i)$.
Since there are at most $n$ timestamps in the range of $ts$, the computation of $ts_{fit}(T_i)$
terminates.
\end{proof}

Using $ts_{fit}$ we can define the quality "$t$-fitting" for MC-histories, which formalizes
the generalized rule for reverse edges from above.

\begin{define}
\label{zpassend}
Let $H$ be an MC-history with the transactions $\{ T_1, \ldots, T_n \}$, a timestamp function $ts$
and the MC-serialization graph $MCSG$.
$H$ is \emph{$t$-fitting in respect to $ts$} iff
\[\forall i,j \in \{ 1, \ldots, n \}: \big(T_i \rightarrow T_j \in MCSG \ \wedge \\
ts(T_i) < ts(T_j)\big) \Rightarrow ts(T_i) < ts_{fit}(T_j).\]
\end{define}

Unfortunately, $t$-fitting MC-histories with $t$-ordered $rw$-projections don't have to be MC-serializable.
We need two additional qualities to prove a respective theorem.
"Irreflexive" avoids edges $T_i \rightarrow T_i$ in an MC-serializability graph.
For an operation sequence of the kind $w_i[x] < r_k^l[x] < m_j^{k,l}$
"$rm$-ordered" ensures that $ts(T_i) < ts(T_j)$ holds, if $T_i$ and $T_j$ commit.
Luckily, both qualities are uncritical when realizing a corresponding serializability protocol.

\begin{define}
An MC-history $H$ is \emph{irreflexive} iff
\[\exists w_i[x], m_i^{k,l}, r_k^l[x] \in H: r_k^l[x] < w_i[x] < m_i^{k,l} \Rightarrow a_i \in H.\]
\end{define}

Consider a client transaction $T_i$ which causes a write operation $w_i[x]$ at the server.
The base protocol from Section \ref{integration}
causes cached method results to be removed from the client's cache right before the method invocation
causing $w_i[x]$ returns control to the client code. Therefore a cache hit corresponding to $m_i^{k,l}$
with  $w_i[x] < m_i^{k,l}$ cannot happen and the base protocol ascertains implicitly "irreflexive".

\begin{define}
\label{rmzgeordnet}
An MC-history $H$ is \emph{$rm$-ordered} in respect to a timestamp function $ts$ iff
\begin{gather*}
\forall i,j \in \{ 1, \ldots, n \}: (\exists w_i[x], m_j^{k,l}, r_k^l[x] \in H:
w_i[x] < r_k^l[x] < m_j^{k,l}) \Rightarrow \\
\big(a_i \in H^{MC}\ \vee a_j \in H^{MC} \vee ts(T_i) < ts(T_j)\big).
\end{gather*}
\end{define}

As we will see below, an MC-history is implicitly $rm$-ordered if the $m$-scheduler cooperates with
an $rm$-scheduler that applies a strict two-phase lock protocol.
The next theorem forms the basis of an $m$-scheduler's implementation of OCTP.
It expects the $rw$-scheduler to provide $t$-ordered $rw$-histories.

\begin{theorem}
\label{zpassser}
An irreflexive MC-history $H$ which is $t$-fitting and $rm$-ordered in respect to a
timestamp function $ts$ is MC-serializable if $RW(H)$ is $t$-ordered in respect to $ts$.
\end{theorem}

\begin{proof}
Assume $H$'s MC-serialization graph $MCSG$ was cyclic. A cycle in $MCSG$ has at least a length of 2,
because for all disjunctive clauses from Definition \ref{defmcsg}
but the case $r_k^l[x] < w_i[x] < m_i^{k,l}$, $i \neq j$ holds for a corresponding edge $T_i \rightarrow T_j$.
However, the case $r_k^l[x] < w_i[x] < m_i^{k,l}$ is excluded because $H$ is irreflexive.
A cycle (with two or more nodes) in $MCSG$ consists of at least one reverse edge.
Otherwise one would obtain a cycle $T_k \rightarrow \ldots \rightarrow T_k$ with normal edges only and so
$ts(T_k) < ts(T_k)$ would hold (contradiction).

The following considerations reveal that for a reverse edge $T_i \rightarrow T_j$
one has got operations $r_k^l[x] < w_j[x]$ and $m_i^{k,l}$ with $ts(T_j) < ts(T_i)$ from the second
disjunctive clause of Definition \ref{defmcsg}.
Edges from the first disjunctive clause of Definition \ref{defmcsg} cannot be reverse edges
because the related operations must not be method operations, but $RW(H)$ is expected to be $t$-ordered.
If an edge from the third disjunctive clause of Definition \ref{defmcsg}
was a reverse edge, then one would have operations $w_i[x] < r_k^l[x] < m_j^{k,l}$
with $ts(T_j) < ts(T_i)$. Yet, this contradicts $H$'s quality to be $rm$-ordered.

Now, let $C$ be a cycle in $MCSG$ and $T_k$ be the node in $C$ with the smallest timestamp.
There must be a reverse edge $T_h \rightarrow T_k \in C$ for some $T_h$ because otherwise $T_k$'s
timestamp would not be minimal in respect to $C$.
Further, let $T_j \rightarrow \ldots \rightarrow T_k$ be the longest acyclic path in $C$
consisting entirely of reverse edges. Then, there must be an edge $T_i \rightarrow T_j \in C$
which is a normal edge. Otherwise $C$ would consist of reverse edges \emph{only} and one would obtain
$C = T_k \rightarrow \ldots \rightarrow T_k$ with $ts(T_k) < ts(T_k)$ (contradiction).

Since $T_i \rightarrow T_j$ is not a reverse edge, one has got $ts(T_i) < ts(T_j)$ and even
$ts(T_i) < ts_{fit}(T_j)$, due to $H$ being $t$-fitting.
Since $T_j \rightarrow \ldots \rightarrow T_k$ only consists of reverse edges, an inductive application
of Definition \ref{zpstempel} results in $ts_{fit}(T_j) \leq ts(T_k)$.
This leads to $ts(T_i) < ts(T_k)$ and contradicts the assumption that $T_k$'s timestamp is minimal in $C$. Thus $MCSG$ must
be acyclic.
\end{proof}

\subsection{Implementation}
\label{impl}

\lstset{numbers=left,numberstyle=\fontsize{7}{7}\selectfont,
basicstyle=\fontsize{8}{8}\selectfont\ttfamily,
commentstyle=\fontsize{7}{8}\selectfont\rmfamily,morekeywords=[1]{each},
columns=fullflexible,keywordstyle=,language=Java,texcl=true,showstringspaces=false}
\begin{figure}[!tb]
\begin{lstlisting}[mathescape]
interface DE {} // Representation of a data element (just a marker interface)
class MId { int k,l; } // ID of a stored method result of operations $r_k^l[x], r_k^l[y], \ldots$
class Op { boolean read; DE x; }
class T { // Representation of a transaction $T_i$
  int id; // The transaction's ID
  List<Op> l = $\emptyset$; int nextMId = 0; // From Figure 4
  int ts   = $\infty$, ts$_{fit}$= $\infty$; // Timestamp and fitting timestamp for $T_i$
  int ts$_{tol}$ = 0; // Maximum timestamp of transactions producing normal edges to $T_i$
  Set<DE> rl = $\emptyset$, wl = $\emptyset$;  // For storing data elements which are read respectively written by $T_i$
  Set<MId> ml = $\emptyset$; // For storing $T_i$'s method operations as \verb|MId|-objects
}

class MScheduler { // Representation of the $m$-scheduler
  int nextTs  = 1;   // To create the next timestamp $ts$
  Rel<DE,MId> V = $\emptyset$;   // Relates $x$ with tuples $(k,l)$ with $x \in d(m_i^{k,l})$
  Rel<DE,T> rt = $\emptyset$; // Relates $x$ with \verb|T|-objects representing $T_i$s such that $r_i^k[x] \in T_i$
  Rel<DE,T> wt = $\emptyset$; // Relates $x$ with \verb|T|-objects representing $T_i$s such that $w_i[x] \in T_i$
  Rel<MId,DE> mt = $\emptyset$; // Relates $(k,l)$ with \verb|T|-objects representing $T_i$s such that $m_i^{k,l} \in T_i$
  Map<int,int> txId2Ts = $\emptyset$; // Relates a transaction's ID with its timestamp
  synchronized void read(T t, DE x, int k) { // Perform $r_i^k[x]$ with \verb|t.id|$=i$
    for each s $\in$ wt(x) if (checkTimestamps(s, t)) { abort(t); return; } // Handle $rw$-conflicts
    t.rl.add(x); rt.add(x,t); // Update relations
    txId2Ts.put(t.id, $\infty$); t.l.add(new Op(true, x));
  }
  synchronized void write(T t, DE x) { // Perform $w_i[x]$ with \verb|t.id|$=i$
    for each s $\in$ wt(x) $\cup$ rt(x)
      if (checkTimestamps(s, t)) { abort(t); return; } // Handle $ww$- and $wr$-conflicts
    for each m $\in$ V(x) // Handle $wm$-conflicts in respect to "$t$-fitting"
      for each s $\in$ mt(m) if (checkTimestamps(s, t)) { abort(t); return; }
    t.wl.add(x); wt.add(x, t); // Update relations
    t.l.add(new Op(true, x));
  }
  synchronized void methodOp(T t, MId m) { // Schedule $m_i^{k,l}$ at the $m$-scheduler with \verb|t.id|$=i$
    for each x $\in$ V$^{-1}$(m) // Handle $mw$-conflicts
      for each s $\in$ wt(x) if (s.ts < $\infty$) {
          // Update \verb|t|'s fitting timestamp if $m_i^{k,l}$ might cause a reverse edge
          if (s.ts > txId2Ts(m.k) && s.ts$_{fit}$ < t.ts$_{fit}$) t.ts$_{fit}$ = s.ts$_{fit}$;
          // If $m_i^{k,l}$ might cause a normal edge, then update \verb|t.ts|$_{tol}$
          if (s.ts <= txId2Ts(m.k) && s.ts > t.ts$_{tol}$) t.ts$_{tol}$ = s.ts;
          if (t.ts$_{tol}$ >= t.ts$_{fit}$) { abort(t); return; } } // Check invariant and abort at a violation
    t.ml.add(m); mt.add(m, t); // Update relations
  }
  synchronized commit(T t) { // Handle commit of \verb|t|
    t.ts = nextTs++; txId2Ts.put(t.id, t.ts); // Create the timestamp
    if (t.ts$_{fit}$ == $\infty$) t.ts$_{fit}$ = t.ts; // Adjust $ts_{fit}$ if necessary
    for each x $\in$ t.wl // Update fitting timestamps for active transactions
      for each m $\in$ V(x)
        for each s $\in$ mt(m) {
          if (s.ts == $\infty$ && t.ts$_{fit}$ < s.ts$_{fit}$) s.ts$_{fit}$ = t.ts$_{fit}$;
          if (s.ts$_{tol}$ >= s.ts$_{fit}$) abort(s); }
    for each x $\in$ t.rl // $rw$-conflicts // Abort transactions violating "$t$-fitting" due to \verb|t|'s timestamp
      for each s $\in$ wt(x) if (checkTimestamps(t, s)) abort(s);
    for each x $\in$ t.wl // $ww$- and $wr$-conflicts
      for each s $\in$ wt(x) $\cup$ rt(x) if (checkTimestamps(t, s)) abort(s);
    for each m $\in$ t.ml // $wm$-conflicts
      for each x $\in$ V$^{-1}$(m)
        for each s $\in$ wt(x) if (checkTimestamps(t, s)) abort(s);
  }
  boolean checkTimestamps(T a, T b) {
      if (a.ts < $\infty$ && b.ts == $\infty$ && a.ts > b.ts$_{tol}$) b.ts$_{tol}$=a.ts;
      return b.ts == $\infty$ && b.ts$_{tol}$ >= b.ts$_{fit}$;
  }
  synchronized void abort(T t) { ... } // Abort \verb|t|
  ...
}
\end{lstlisting}
\caption{Java Pseudo Code for "$t$-fitting" at the $m$-Scheduler}
\label{tsfitcode}
\vspace{-1cm}
\end{figure}

This section characterizes a serializability protocol for an $m$-scheduler
which is derived from Theorem \ref{zpassser}.
We assume that the $rw$-scheduler applies a strict two-phase lock protocol since this
protocol is common for commercial database management
%(we don't care if the protocol 1- or 2-version-based)
systems.\footnote{This is just \emph{some}
legitimate assumption for realizing a respective protocol -- what matters most
is that the $rw$-scheduler produces $t$-ordered $rw$-histories according to a timestamp
function whose ordering is known to the $m$-scheduler.}

As mentioned at the beginning Section \ref{protocol}, in case of a strict two-phase lock protocol,
the commit order of $rw$-transactions may be considered a timestamp order.
More specifically, the timestamp function is implicitly given by $ts(T_i) < ts(T_j) :\Leftrightarrow c_i < c_j$.
(As we will see, aborted transactions are not of interest.)

Since the corresponding $rw$-histories are strict, the situation $w_i[x] < r_k^l[x] < m_j^{k,l}$
leads to $w_i[x] < c_i < r_k^l[x] < m_j^{k,l} < c_j$ and so $ts(T_i) < ts(T_j)$ holds due to the chosen
timestamp function. Hence, the quality "$rm$-ordered" is automatically guaranteed.
For serializability the $m$-scheduler only needs to ensure "$t$-fitting".

Figure \ref{tsfitcode} captures a respective implementation using Java pseudo code and forms an extension
of the base protocol's pseudo code from Figure \ref{baseprotcode}.
For simplicity, it assumes that the $m$-scheduler is notified of transactional operations by calls to
the methods \texttt{read()}, \texttt{write()}, \texttt{commit()}
and \texttt{abort()}. The method \texttt{methodOp()} handles $m$-operations and is called by \texttt{handleReqest()}
from Figure \ref{baseprotcode}.
Except for \texttt{abort()}, the methods do not impact
the systems's normal transaction management process but only observe it. However, a call to
\texttt{abort()} is assumed to abort the client-side transaction
as well as related resource manager transactions.

For the $m$-scheduler to work properly, it is required that an underlying resource manager processes
read, write, commit and abort operations in the same order as they are observed by the $m$-scheduler.
Further, all those operations must pass the $m$-scheduler.
The implementation does not yet account for memory management but in fact, all of the code's data structures can
be handled in a way such that their size remains limited. At the end of this section we will explain
how this can be realized.

%Further, all transaction operations are supposed to pass the $m$-scheduler
%before they reach \texttt{rm}.
%The pseudo code's focus is on the data structures that the
%$m$-scheduler has to maintain in order to assert $t$-fitting transactions:
Transactions are represented by instances of class \texttt{T}, whereby a transaction's timestamp
as well as its fitting timestamp are initially unknown. For that reason, \texttt{T.ts} and
\texttt{T.ts}$_{fit}$ obtain the value $\infty$ when a respective transaction begins (Line 7).
The lists \texttt{rl}, \texttt{wl} and \texttt{ml} (Lines 9, 10) store transaction operations
in order to detect conflicts with other transactions. The lists are used at a transaction's
commit-time in order to find conflicts with active transactions (Lines 46 to 57).

Let $T_i=$\texttt{t} be a transaction which is represented by an instance of \texttt{T}.
The field \texttt{t.ts}$_{tol}$ from Line 8 stores the largest timestamp of a committed transaction producing
a normal edge which points to \texttt{t}. \texttt{t.ts}$_{tol}$ is important to guarantee "$t$-fitting"
throughout \texttt{t}'s lifetime: While normal edges pointing to \texttt{t}
may increase the value of \texttt{t.ts}$_{tol}$,
reverse edges originating from \texttt{t} may decrease \texttt{t.ts}$_{fit}$ dynamically
due to new transactional operations.
The $m$-scheduler's main task is to ascertain \texttt{t.ts}$_{tol}$\texttt{ < t.ts}$_{fit}$ until \texttt{t}
commits. At a violation of this invariant it aborts either \texttt{t} or it aborts the respective conflicting
transaction. The method \texttt{checkTimestamps()} from Line 59 assists in updating \texttt{t.ts}$_{tol}$
accordingly and in checking the stated invariant after the update.
It is used my the methods \texttt{read()}, \texttt{write()} and \texttt{commit()}.

The relation \texttt{V} associates
data elements (instances of class \texttt{DE}) with cached method
calls (Line 15). The latter ones are identified by \texttt{MId}-objects
according to the read operations by which the method result was
computed. (This coincides with the description of $V$ from Section
\ref{integration} and Figure \ref{baseprotcode}.) The purpose of the relations \texttt{rt},
\texttt{wt} and \texttt{mt} is to associate data elements
respectively IDs of method results with transactions in which they
were accessed (Lines 16 to 18).

The methods \texttt{read()}, \texttt{write()} and \texttt{methodOp()}  first check whether the
intended operation might violate the quality "$t$-fitting". At a violation, they abort
the current transaction. (Note that the pseudo code abstracts from the details
of the abort process.) Otherwise, they update
the $m$-scheduler's data structures.

As an example of how the violation check works, consider the Line
21 of \texttt{read()}: Using \texttt{wt} the method binds
each transaction that wrote the same data element as the current
read operation to the local variable \texttt{s}. If the
transaction (bound to) \texttt{s} has got a timestamp less than
$\infty$ it must have committed and so if the current transaction
\texttt{t} committed too, the read operation would result in a
normal edge \texttt{s}$\rightarrow$\texttt{t}$ \in MCSG$. So, in
order to assert "$t$-fitting" for \texttt{t} the expression
\texttt{s.ts < t.ts}$_{fit}$ must hold and this is just checked in Line 21 using \texttt{checkTimestamps()}.
The arguments behind the checks of the method \texttt{write()} are similar (Lines 26 to 29).

\texttt{methodOp()} observes a new $m$-operation $m_i^{k,l}$ of a transaction $T_i=$\texttt{t} and determines
if the operation produces reverse or normal edges in respect to committed transactions.
In order to do so, \texttt{methodOp()} loops over all data elements which are referenced
by the $m$-operation $m_i^{k,l}$ (Line 34). If a committed transaction $T_j=$\texttt{s} has written one of those
data elements, there is a conflict between $T_i$ and $T_j$. Further, if $T_j$'s
timestamp is younger than $ts(T_k)$, one obtains the situation $r_k^l[x] < c_k < w_j[x] < c_j < m_i^{k,l}$
which implies a reverse edge and so \texttt{t.ts}$_{fit}$ must potentially be updated (Line 37).
Using the map \texttt{txId2Ts} the $m$-scheduler fetches the timestamp $ts(T_k)$ in respect to $m_i^{k,l}$.
Conversely, $ts(T_j) \leq ts(T_k)$ only allows the two options
$w_j[x] < r_k^l[x] < m_i^{k,l}$ and $r_k^l[x] < w_k[x] < c_k < m_i^{k,l}$ with $k = j$.
The former option indeed causes a normal edge and so, \texttt{t.t}$_{tol}$ must potentially be updated (Line 39).
The latter option is
impossible since the base protocol causes the cached result referenced by $m_i^{k,l}$
to be invalidated right after executing of $w_k[x]$.
Eventually, \texttt{methodOp()} tests the above stated invariant for \texttt{t} due to the potential change of
\texttt{t.ts}$_{fit}$ and \texttt{t.t}$_{tol}$ (Line 40).

Finally consider the functioning of
\texttt{commit()}: At first a
timestamp is assigned to \texttt{t.ts} (Line 44). Since
\texttt{commit()} is synchronized, all committing transactions are
totally ordered and so is their timestamp. In concordance with
Definition \ref{zpstempel} \texttt{t}'s fitting timestamp is set
to \texttt{t.ts} if it hasn't got a lower timestamp yet (Line 45).

Because now, \texttt{t}'s timestamp is known,
all conflict edges between \texttt{t} and active transactions can be
checked to see whether they are reverse or normal edges and if they violate
"$t$-fitting".
The Lines 46 to 49 determine all related reverse edges and update
an active transaction's fitting timestamp \texttt{s.ts}$_{fit}$
accordingly. Note that a related conflict is guaranteed to cause a reverse edge.
To see this, let again be $T_i = $\texttt{t} and $T_j = \texttt{s}$. A normal edge would lead
to the situation $w_i[x] < r_k^l[x] < m_j^{k,l} < c_i$ but this contradicts the assumption
that the resource manager guarantees strictness for $rw$-histories.
Line 50 checks if \texttt{s} must be aborted because of a change of \texttt{s.ts}$_{fit}$ in Line 49.

The Lines 51 to 57 inspect active transactions \texttt{s} for normal edges
\texttt{t}$\rightarrow$\texttt{s}$ \in MCSG$ and abort a respective transaction
\texttt{s} if $t$-fitting is violated due to \texttt{t}.
In analogy to the case from the Lines 46 to 49, it can be shown that conflicts
inspected by the Lines 55 to 57 always lead to normal edges.

\subsection{Memory Management}

So far the data structures used in Figure \ref{tsfitcode} would unboundedly grow
with the number of transactions and operations that the system processes.
The following paragraphs briefly describe how to limit the size of these data structures
without changing the functioning of the discussed implementation.

The first question to answer is when entries for a certain transaction
may be deleted because they don't affect the processing of active transactions anymore.
A closer look at Figure \ref{tsfitcode} leads to two different cases to be considered:
Due to the Lines 21, 27, 29, 40, 52, 54, and 57 an (active) transaction $T_i$
may be aborted if some other transaction $T_j$ produces a normal edge $T_j \rightarrow T_i$ such that
$ts(T_j) \geq ts_{fit}(T_i)$ holds. For this case it suffices to retain the entries for
just those transactions contained in the following set:
\begin{center}
$M_1 = \big\{$ \texttt{t}$\ |\ $\texttt{t.ts} $\geq \min \{\ $\texttt{s.ts}$_{fit}\ |\ $\texttt{s} is active $\}\big\}$.
\end{center}

The second case covers Line 37 where the fitting timestamp of a committed transaction
is assigned to the fitting timestamp of an active transaction. Therefore,
one also needs to retain the entries of transactions \texttt{t} contained in the following set:
\begin{center}
$M_2 = \big\{$ \texttt{t}$\ |\ $\texttt{t.ts} $\geq \min \{\ $\texttt{s.ts}$_{fit} | \exists$\texttt{(x,(k,l))}$\in$\texttt{V}$:$
\texttt{x} $\in$ \texttt{s.wl}$\ \wedge\ $\texttt{txId2Ts(k)} $ < $ \texttt{s.ts}$\}\big\}$.
\end{center}
Finally Line 49 also affects the fitting timestamp of active transactions but since it
only passes on the fitting timestamp of a transaction that is about be committed, the respective entry
is already contained in $M_1$.

The joint set $M_1 \cup M_2$ forms the set of transactions whose entries need to be retained, but how can its size
be controlled? There are two ways to do this: Firstly, one can delete entries \texttt{(x,(k,l))} from \texttt{V}
which are stale because some transaction $T_j$ with a younger timestamp than $T_k$
has preformed an operation $w_j[x]$. This reduces the size of $M_2$.
Alternatively, an active transaction can be aborted in order to reduce the size of $M_1$.
Finding the right candidates to be removed from $M_1 \cup M_2$ can be done efficiently.
(A detailed discussion of this process is beyond the scope of this paper.)
Moreover, practical experience such as from the experiments of the next section show
that the size of $M_1 \cup M_2$ is not a critical system factor.

By controlling $|M_1 \cup M_2|$, one can limit the size of the data structures \texttt{rt}, \texttt{wt}, \texttt{mt} and
\texttt{txId2ts} from Figure \ref{tsfitcode}. Still, \texttt{V} may grow unboundedly because it must hold an entry
for every valid cached method result but there may be arbitrary many of those results (in arbitrary many caches).
To tackle this problem, \texttt{V} should be limited by a fixed (but reasonably high) upper bound. Then, an
LRU-strategy can be used to replace respective entries in \texttt{V}. By extending the
base protocol from Section \ref{integration} the client cache
that stores a method result which is associated with a replaced entry of \texttt{V}
can be notified in order to erase the result.

A last thing to consider is that due to invalidation delays for cached method results, \texttt{methodOp()} can potentially be called with
an argument value \texttt{m}
for which the respective entry in \texttt{V} has already been replaced (or removed by controlling $M_2$).
For this reason \texttt{methodOp()} must be adjusted to check the validity its argument value \texttt{m}. To do so, the following code
should be inserted after Line 33 of Figure \ref{tsfitcode}:
\begin{center}
\texttt{if (V}$^{-1}$\texttt{(m) = }$\emptyset$\verb|) { abort(t); return; }|
\end{center}

\section{Evaluation}
\label{evaluation}

In this section we briefly justify the intellectual investment in transactional method caching
by giving evidence that the approach can considerably improve system scalability and performance.
%For a lack of space, the related experiments are just presented to the extend needed to suit this
%purpose.

\subsection{Experiment}

We implemented a prototype of a transactional method cache and an $m$-scheduler on top of
the EJB application server product JBOSS v3.2.3 \cite{jboss}. The implementation of the cache's base
protocol follows the architecture from Section \ref{integration}. The relational database management system
MySQL v4.0.18 \cite{mysql} serves as a resource manager. The client is a multithreaded Java program
performing remote service method invocations. The client, the application server
and the database system are hosted on three separate PCs in a local network, whereby the PCs'
hardware suits up-to-date desktop standards (including a 1.2 GHz Pentium 4 Processor and 512 MB RAM).
The PCs operate under Windows XP.
By observing the related system resources we ensured that neither network bandwidth
nor the load on the client machine represented a potential bottleneck for the experiment.

The experiment's database consists of a single SQL table with the
following structure:
\begin{verbatim}
item(id int primary key, name varchar(50), descr varchar(250),
     price float, weight float, manuf varchar(50))
\end{verbatim}
Using an auxiliary program the table was filled
with 1 million random valued entries. At the application server,
an EJB session bean implemented a service interface according to
Figure \ref{rwinterface}. The method \texttt{findItemById()} reads
a database entry from the \texttt{item}-table via JDBC \cite{jdbc}
and returns the contents of a related table row as an
\texttt{Item}-object. The related table row is queried via its key
value using the method's \verb|id|-argument. Similarly
\texttt{updateItem()}, changes a table row according to the
\verb|Item|-object which is passed in as an argument. The related
table row is accessed via its key value using the
\texttt{Item}-object's \texttt{id}-field. (If no such row
exists, the method throws an exception.) For the database the SQL
isolation level was set to "SERIALIZABLE". On this level, MySQL
performs a strict (1-version)
%2-version
two-phase lock protocol with row level locking.

The $m$-scheduler is implemented as a delegating JDBC driver and incorporates
the protocol from Section \ref{impl}. As explained at the end of Section \ref{integration},
we had to insert extra code behind the service methods' JDBC statements
in order to inform the $m$-scheduler about the accessed table rows.
(In this respect, the corresponding \texttt{id}-value was chosen to identify a data element.)

%Moreover, we implemented another serializability
%protocol for the $m$-scheduler which has not been discussed. In essence this second
%serializability protocol is similar to the classical OCC protocol such as known from \cite{}.
%Since OCC is known for causing a high transaction abortion rate at high transaction contention,
%it useful for
The client contains a single transactional method cache. The cache applies an
LRU replacement strategy with a limit of 4000 storable method results.
A variable number of client threads perform transactions concurrently.
Every transaction consists of 10 method calls addressing the server's EJB interface.

For every call a client thread chooses randomly whether to call \texttt{findItemById()} or \texttt{updateItem()}.
\texttt{findItemById()} is invoked with the probability $p_r = 0.8$ whereas \texttt{updateItem()} has the probability $1 - p_r$.
After finishing the 10 calls successfully, the thread commits (respectively aborts)
its transaction with a chance of $p_c = 0.95$ (respectively $1 - p_c$).\footnote{We have also tried other transaction lengths
varying between 5 and 25 calls per transaction. The results are very similar to the chosen value of 10 method calls per transaction.}
At last the thread pauses for 1 second before starting a new transaction
(no matter if the previous transaction committed or aborted).

\lstset{numbers=left,numberstyle=\fontsize{8}{9}\selectfont,
basicstyle=\fontsize{8}{9}\selectfont\ttfamily,
commentstyle=\fontsize{8}{9}\selectfont\rmfamily,morekeywords=[1]{each},
columns=fullflexible,keywordstyle=,language=Java,texcl=true,showstringspaces=false}
\begin{figure}[!tb]
\begin{lstlisting}
public interface ItemSession extends javax.ejb.EJBObject {
   public Item findItemById(int id) throws RemoteException;
   public void updateItem(Item item) throws RemoteException;
}

public class Item implements java.io.Serializable {
  public int id; public String name;
  public String description; public double price;
  public double weight; public String manufacturer;
}
\end{lstlisting}
\caption{Java Pseudo Code of the Experiment's Service Interface}
\label{rwinterface}
\end{figure}

An important parameter that determines the experiment's cache hit rate as well as the cache
invalidation rate is the value of the \texttt{id}-argument when
calling \texttt{findItemById()} and the value of \texttt{item.id} when
calling \texttt{updateItem()}. The client uses a random distribution to compute a corresponding value,
whereby 1 million \texttt{item}-table rows are potentially referenced.

During a warmup phase the cache fills up to its maximum size of 4000 method results.
After that the probability that a service method call causes a hit is
$53\%$ (this chance implies the event of invoking \texttt{findItemById()}). The probability is mainly
caused by the given cache size and the chosen random distribution for generating \texttt{id}-values which is not uniform.\footnote{
Essentially we employed a log-normal distribution with the standard parameters $\mu = 7$ and $\sigma = 1.6$.}
The chance of invalidating a cached method result (due to a respective call of
\texttt{updateItem()}) is about $13.25\%$ ($= (1-p_r) / p_r \cdot 53 \%$).

One may ask, why we did not resort to an existing benchmark application instead of
designing the experiment from above. Unfortunately there are no useful and
realistic benchmarks for testing client-side transactions in the application server domain.
RUBiS \cite{rubis1,rubis2,rubislink} is an EJB-benchmark that comes close to our needs and models
an auction web site which is similar to eBay.com. However, the benchmark does not
account for client-side transactions and cannot be reasonably adjusted to make use of this
feature.

Still, the main input parameters that govern the experiment from above represent conservative estimates
of similar parameters that result from applying \emph{non-transactional method-caching} to RUBiS.
In particular, \cite{pfeifer2} observed cache hit rates between 53\% and 78\% when applying non-transactional
method caching to RUBiS. \cite{rubis2} considers a fraction of about 85\% of read-only method calls as most representative
for an auction web site workload. (In contrast, we are more conservative by setting $p_r = 80\%$.)

We therefore believe, that transactional method caching can cause similar results as for the given experiment when
it is applied to real world applications.
Moreover, due to the experiment's simplicity, its input parameters
are clear and its results are well traceable.
Beyond these considerations, \cite{pfeifer2} has already shown that
non-transactional method caching produces very good efficiency improvements when applied to RUBiS.

\subsection{Results}

\begin{figure}[!tb]
%\begin{center}
\centering
\includegraphics[width=11cm]{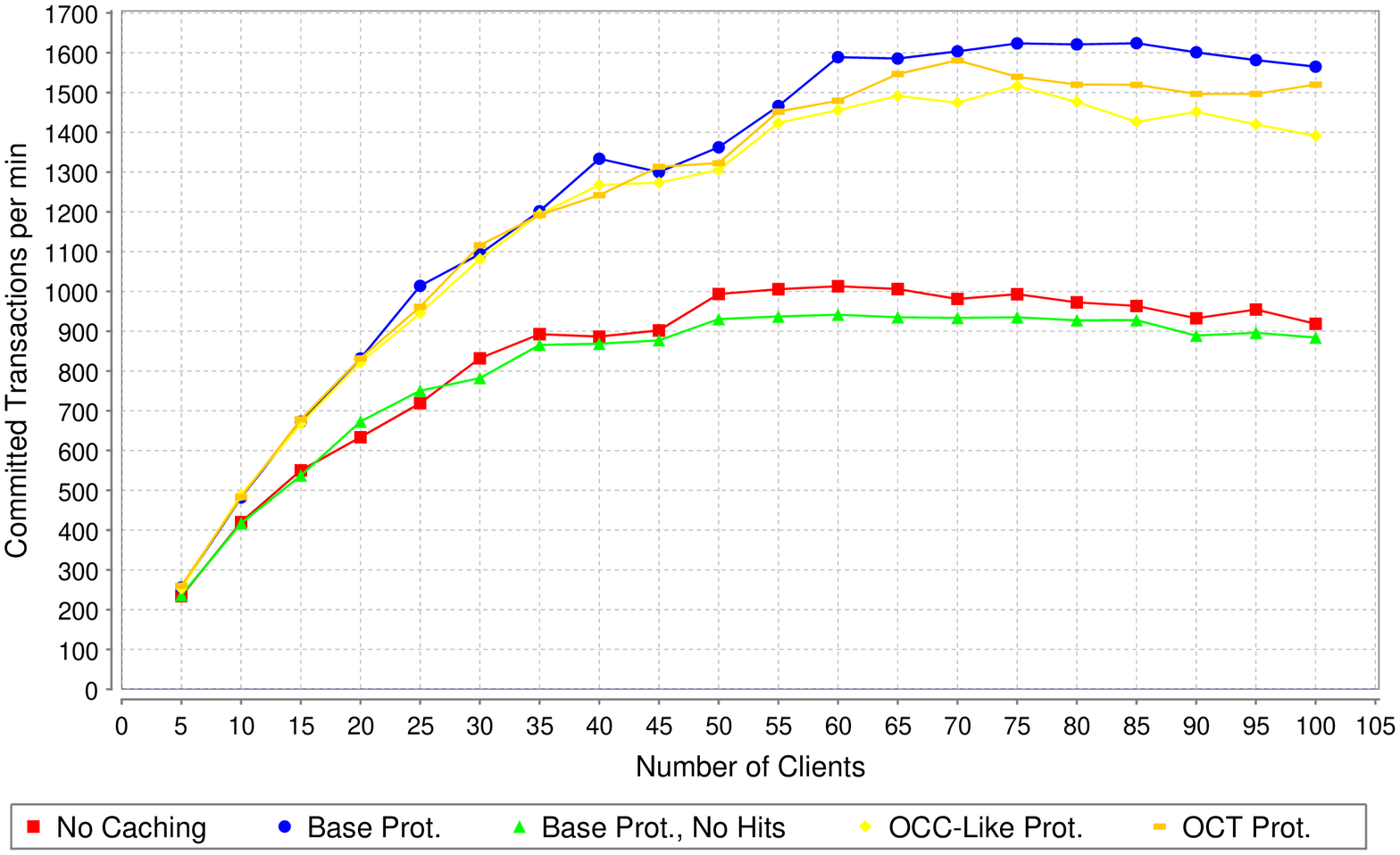}
%\end{center}
\caption{Committed Transactions as a Function of the Number of Concurrent Client Threads (Throughput)}
\label{txscale}
\end{figure}

\begin{figure}[!tb]
\centering
\includegraphics[width=11cm]{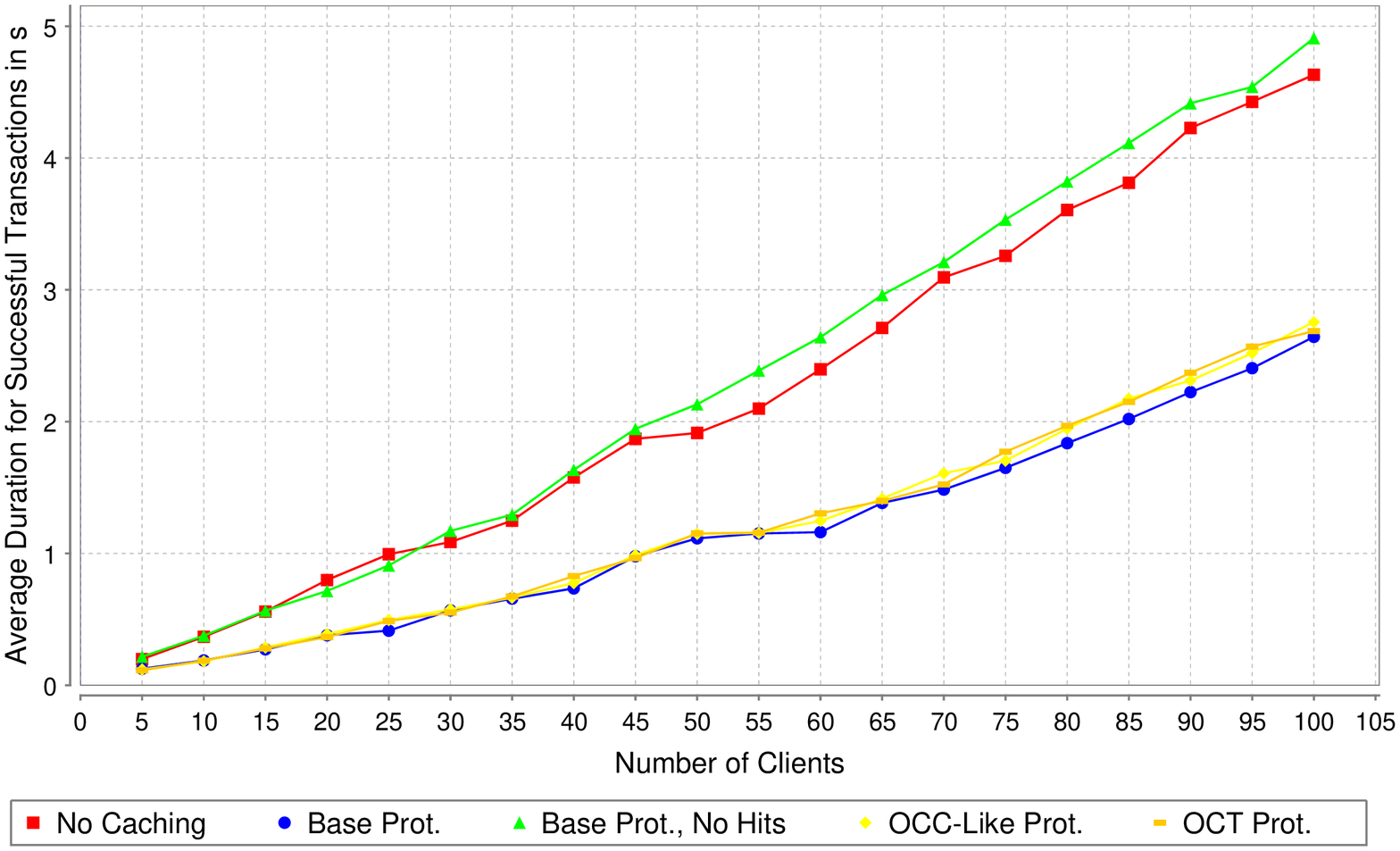}
\caption{Average Duration of a Transaction that Executed 10 Service Method Calls (Response Time)}
\label{txperf}
\end{figure}

\begin{figure}[!tb]
\centering
\includegraphics[width=11cm]{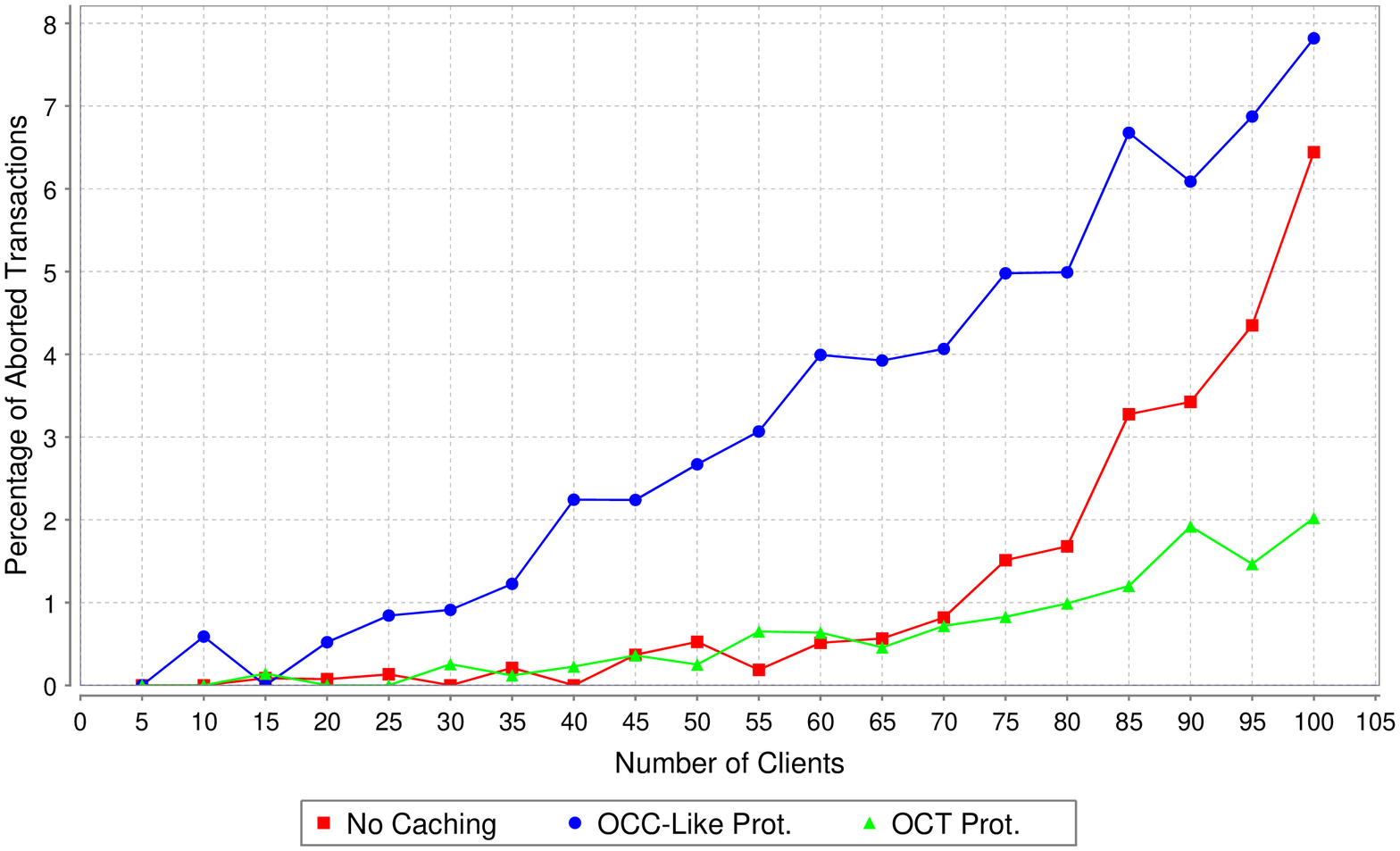}
\caption{Percentage of Aborted Transactions in Respect to Started Transactions}
\label{txabort}
\end{figure}

For the results presented next, every data point corresponds to a two minute measuring period.
The measuring period was preceded by a two minute warmup phase in order to fill the method cache.
By conducting additional test experiments we ensured that both the duration of the measuring phase
as well as the warmup phase produced representative values.

Figure \ref{txscale} shows the number of committed transactions per minute for a varying number of concurrent
client threads under five different system configurations. The graph "No Caching" represents the
respective results for the system without using a method cache. The graph "OCT Prot." depicts the results
if transactional method caching is applied using the $m$-scheduler protocol from
Section \ref{protocol} (OCTP). A simpler transactional protocol which is similar to the classical
OCC protocol from \cite{223787} has also been tested (see also Section \ref{protocolformalism}).
The fourth graph displays system behavior when a method cache is used while \emph{only} applying the
base protocol from Section \ref{baseprot}. This option would hardly be applied in practice since
it does not provide transactional consistency. It was added to Figure \ref{txscale} because
it gives an impression of the overhead of an $m$-scheduler protocol as opposed to the pure base protocol.
Similarly the graph "Base Prot., No Hits" shows system behavior when applying the base protocol
but not granting any cache hits. This graph helps to characterize the overhead of the base protocol
versus a system without method caching.

All system variants scale well with an increasing number of concurrent client threads.
However, system variants using method caching attain a considerably higher
level of transaction throughput. By comparing "No Caching" and "Base Prot., No Hits" one can see
that the additional cost for the base protocol remains moderate.
The $m$-scheduler protocols reduce the transactional throughput in comparison to a "pure" base protocol, because
they abort a fraction of transactions for consistency reasons.

Figure \ref{txperf} illustrates the average duration of a successful transaction for the same runs
as in Figure \ref{txscale}. Here, method caching considerably shortens transaction runtimes and so it
improves system performance. As in Figure \ref{txscale} one can observe the cost of the base protocol and the
$m$-scheduler protocols which are both moderate.

Finally, Figure \ref{txabort} shows the transaction abortion rate for those runs from Figure
\ref{txscale} which maintain transactional consistency. Obviously abortions become more likely with
an increasing number of concurrent transactions.
The worst abortion rate is observed for the OCC-like protocol -- transactions may be aborted
by the $m$-scheduler as well as the database system. For the system variant without method caching
only the database system aborts transactions. Surprisingly a system with method caching using OCTP
has lower abortion rates than the variant without method caching! The reason for this is that
OCTP allows even (some) transactions to commit that have caused cache hits on stale cached method results.
As opposed to that, the OCC-like protocol always aborts transactions accessing stale cached method results and
so, OCTP has a better quality. In essence, OCTP establishes a kind of a consistent
multi-version transaction scheduling policy in respect to cached method results.

All in all, the experiments give evidence that \emph{using OCTP,
transactional method caching can improve system throughput, response time as well as transaction abortion rates}.

\section{Related Work}
\label{relatedwork}

\subsection{Web Application Caching}

In the last years, research as well as industry has made
various efforts to improve the performance of web applications by
means of caching. Since transactional method caching can be beneficial in the
context of web applications, we briefly compare it against other caching approaches
in this field and discuss the advantages and disadvantages.

\begin{figure}[!tb]
\centering
\includegraphics[width=3.5cm]{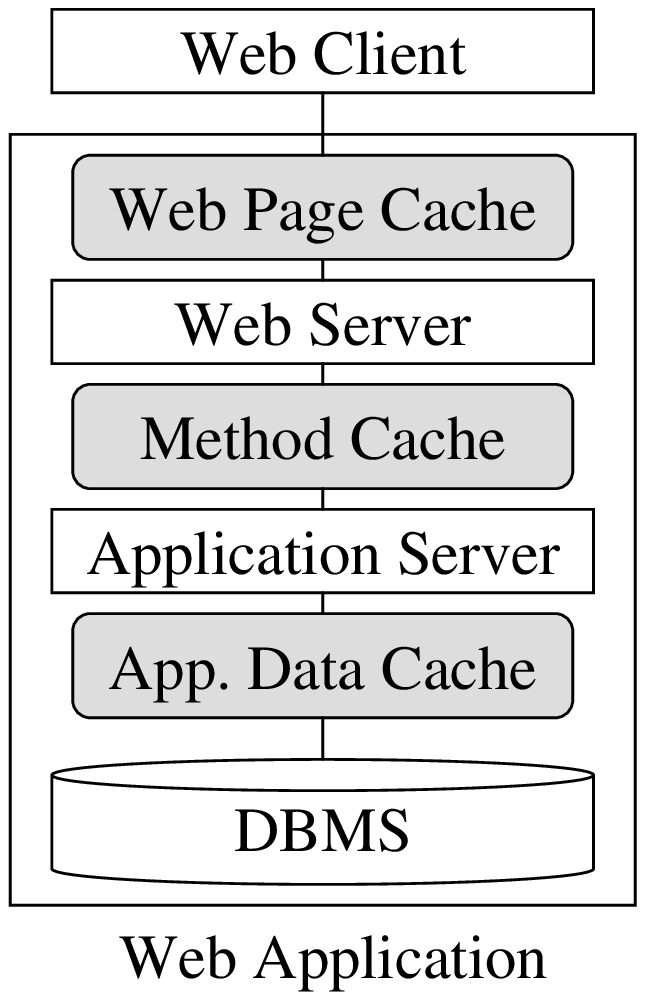}
\caption{Common Tiers of Web Application Architectures and Related
Options for Caching.} \label{webapplayers}
\end{figure}

Figure \ref{webapplayers} shows the tiers of a typical web
application architecture and highlights where caches potentially
come into play:
\begin{itemize}
\item\ Application data caching happens somewhere in between the database and the application
server tier. If it is done right in front of the database
\cite{oracledbcache,db2,sql-serverdbcache,timesten_2},
abstractions of database queries are associated with query results
in the cache. In case of a cache hit, the query result is
immediately returned by the cache as opposed to running the
database query engine. At the server side, application
data is cached either programmatically through runtime objects
whose structure has been designed by the application developer
\cite{jcs,jcache} or it is controlled by an
object-relational mapping framework \cite{toplink,jdx}.

\item\ Web page caching usually occurs in front of a servlet- or script-enabled web server.
Beyond the simple task of caching static pages, there are also
many approaches for caching dynamically generated web pages
\cite{oracle_reverse_proxy,challenger99scalable,cache_portal_2}.

\item\ A method cache is inserted at the "backend" of a servlet- or script-enabled web server
from where application server calls are initiated. While \cite{pfeifer2} discussed non-transactional
method caching, this paper is the first one presenting a solution for transactional method caching.
\end{itemize}

The major problem of application data caches is that they can only save the cost
of database queries but no cost originating at the application server tier.
Therefore caching of service method results has a higher potential for improving system efficiency.
In contrast, \emph{the pure cost for executing page generation scripts at the Web server tier is rather low and so,
there is not much gain when caching dynamic Web pages instead of service method results.}

One important question that all dynamic web caching strategies must
deal with is when and how to invalidate cache content.
In \cite{cache_portal_1,cache_portal_2} and
\cite{form_based} URLs of dynamic pages on the web server side are
associated with dependent SQL queries on the database level. If a
database change affects a corresponding query, the related pages
in the cache are invalidated.
In \cite{cache_portal_1,cache_portal_2} dependencies
between queries and URLs are automatically detected through
sniffing along the communication paths of a web application's
tiers. Although the approach observes database changes,
\emph{it provides only a weak form of update consistency, whereas
our approach ascertains full transactional consistency.}

Other strategies for dynamic web page caching require a
developer to provide explicit dependencies between URLs of pages
to be cached and URLs of other pages that invalidate the cached
ones \cite{dynamai}. Often, server-side page generation scripts or database systems may
also invalidate a cached page by invoking invalidation functions
of the web cache's API \cite{oracle_reverse_proxy,spidercache,xcache}.
Unfortunately these strategies are
invasive which means that application code (e.g. page generation
scripts) has to be changed. In contrast, \emph{our approach is completely
transparent to the client code and requires only minor changes at
the server-side code.} Therefore it can be applied even in late
cycles of application development.

An explicit fragmentation of dynamic web pages via annotations in
page generation scripts helps to separate static or less dynamic
aspects of a page from parts that change more frequently
\cite{comparison,esi}. Also, dependencies such as
described in the previous paragraph can then be applied to page
fragments instead of entire pages. In this respect, \emph{our approach
enables an even more fined grained fragmentation as it treats
dependencies on a level where page scripts invoke service methods
from the application server.} A great benefit, is that explicit
page fragmentation annotations (such as supported by \cite{esi})
then become obsolete. This also leads to the conclusion that
\emph{caching the results of service method calls causes cache hit rates which
are at least as good as in the case of dynamic Web caching (or even better).}

\subsection{Conventional Transaction Protocols}
\label{convproto}

This section highlights the differences between conventional transaction protocols
and the approach described in this paper.
%In particular it explains
%why transactional method caching is not just a special case of conventional transactional caching.

Existing work in the field of transactional caching relates to page server
systems, where a client can download a database page to its local cache, change it and
eventually send those changes back to the server \cite{franklin97transactional}. For these systems the cache protocol ensuring transactional
consistency forms an integral part of the database system itself. In contrast, this paper's approach assumes that
a tight integration with a given database system is not possible. Moreover, the presented approach accounts for the characteristics
of an application server that does not enable direct access to data elements such as pages.
Therefore, we described how to extend an application server architecture to enable consistent client-side method caching.
The cache protocol is designed so that is does not alter the standard communication flow
between client and server. Also, the unit for ensuring the transactional consistency
-- the $m$-scheduler -- remains separate from an underlying resource manager (such as a database system).

In order to develop an efficient protocol for the $m$-scheduler we presented a theory for reflecting
the use of cached method results inside transactions. Without this theory, proving the correctness
of OCTP would have been very difficult. As opposed to that, the correctness of conventional
transactional cache protocols such as OCC \cite{223787} or CBR \cite{franklin97transactional} is more obvious and does not
demand formal considerations.

An important difference between OCTP and other conventional transactional cache protocols
is that OCTP does neither avoid access to stale cache entries (such as CBR)
nor necessarily abort transactions which have accessed stale cache entries (such as OCC).
Therefore, in spite of being optimistic, OCTP can offer low transaction abortion rates.

With respect to the taxonomy of \cite{franklin97transactional} OCTP is a "detection based protocol" whereby a validation may be "deferred until commit".
Further, OCTP gives invalidation hints "during a transaction" and uses "invalidation" (as opposed to "propagation") as its "remote update action".
Propagation as a remote update action is not applicable since the $m$-scheduler has no access to a
method call's arguments which are needed for recomputing the method result that would have to be propagated.
According to the taxonomy of \cite{gruber} OCTP supports "early aborts" and may be classified as "lazy reactive".

Apart from transactional cache protocols, OCTP has a similarity to the multiversion timestamp protocol (MVTO)
from \cite{357355}. Let $T_i$ be a transaction with an operation $r_i[x]$ but without a prior $w_i[x]$.
At MVTO, $r_i[x]$ reads the version $x_k$ that was written by a \emph{committed}
transaction $T_k$ such that
$ts(T_k) = \max \{ ts(T_j)\ |\ ts(T_j) < ts(T_i) \wedge w_j[x] \in T_j \}$ holds.
Scheduling an operation $m_i^{k,l}$ at OCTP is similar to scheduling $r_i[x]$ at MVTO.
However, at OCTP the version of a respective data element is already fixed by the cache hit itself, namely by $m_i^{k,l}$.
Therefore, the $m$-scheduler cannot choose $x_k$ but can only determine where $T_i$ would best "fit" in the given timestamp order.
In order to do so the $m$-scheduler computes $ts_{fit}(T_i)$.

The fitting timestamp $ts_{fit}$ from Definition \ref{zpstempel} is also connected to the concept of dynamic timestamps from \cite{bayer}.
In \cite{bayer} a scheduler may delay the assignment of timestamps to transactions in order to accept a broader range
of serializable histories.
A respective timestamp is therefore called \emph{dynamic}.
Although OCTP's fitting timestamp may change dynamically, a related transaction's real timestamp $ts(T_i)$
is always dictated by the $rw$-scheduler and therefore it is not dynamic. This is the crucial difference
between OCTP and the proposition from \cite{bayer}.

\section{Conclusion}
\label{conclusion}

This paper has presented an approach for the transactional caching of method results in
the context of application server systems. A related cache is placed at the system's client
side. It comes into play when the client performs a sequence of method calls addressing the server,
whereby the calls are demarcated by an ACID transaction.
If the client invokes a read only method with the same arguments for the second time the
related result can potentially be taken from the cache which avoids an execution at the server.
For a reasonable hit rate, the approach is inter-transactional meaning that a cached method result
can be used by multiple client transactions.

The paper has adjusted the conventional architecture
of an application server in order to enable transactional method caching.
Since the use of cached method results alters the way a transaction is processed, it must be regarded
when ensuring transactional consistency. Therefore, we introduced an new system component
at the server side which maintains transactional consistency in the presence of cache hits.
This so called $m$-scheduler observes cache hit operations as well
as normal data access operations ascertains serializability of client transactions.

To develop a protocol for an $m$-scheduler, the paper extended the conventional 1-version transaction
theory by an operation which reflects the use of cached method results.
We derived a definition for serializability in respect to the extended transaction histories and proved
a corresponding serializability theorem.

Using these theoretical results, we developed an efficient recovery protocol as well as an efficient serializability protocol for
an $m$-scheduler and proved their correctness. Moreover, the paper discussed some of the protocols' implementation aspects.
An experimental evaluation showed that the presented cache can considerably improve system performance and scalability
as well as transaction abortion rates.

A limitation of the approach is that
in order to guarantee transactional consistency, the $m$-scheduler needs to observe \emph{all} data access operations addressing
an underlying resource manager. Also, it does have to make some basic assumptions
about the resource managers' transaction management protocols.
The clear advantage of having a separate scheduler for $m$-operations is that
an integration of the presented cache only requires the modification of the application server and its clients but not the resource manager(s).
This is especially important for practical considerations since many modern application servers systems are open
and well extendable, but most database management systems are not.

Note, that the stated limitation would be uncritical, if an $m$-scheduler was integrated in a resource manager.
Still, the major contributions of this paper, namely the presented theory, the recovery protocol
and the transactional cache protocol also apply to this case.

As part of our future work we would like to apply the idea behind OCTP to the domain of page server systems.
In this field many transactional cache protocols have been studied (for an up-to-date comparison see \cite{977122}).
However, as explained in Section \ref{convproto},
none of them allow transactions to commit who have accessed stale cache entries.
Currently, transaction protocols for page servers either enable moderate efficiency combined with
low abortion rates (e.g. CBR) or high efficiency combined with potentially intolerable abortion rates (e.g. OCC).
In contrast, an OCTP-like protocol for page servers could bring together
high efficiency (via optimism) and low abortion rates (by tolerating access to stale cache entries) while still ensuring
serializability.

\section*{Acknowledgement}
The authors would like to thank Prof. Birgitta König-Ries for thoroughly proof-reading this paper.

%
% For practical considerations we assumed that $m$-scheduler is designed as a component
% which is integrated in an application server and which is separate from an underlying resource manager, e.g. a database system.
% With the limitations mentioned above it follows that application server should have exclusive access to the database system.
% To make the $m$-scheduler observe data access operations as required, service method implementations
% must be extended by a piece of code that notifies the $m$-scheduler of the data elements accessed during
% a corresponding method execution.
%
% Note that none of the mentioned limitations hold if the $m$-scheduler is integrated at the resource manager
% instead of the application server. However, commercial database systems are not tailored to such an integration.

\bibliographystyle{acmtrans}
\bibliography{literature_clean}

\end{document}